\documentclass[12pt]{iopart}

\usepackage{amssymb}
\usepackage{amsthm}
\usepackage{graphicx}
\usepackage{caption}
\usepackage{subcaption}
\usepackage{comment}
 \usepackage{amssymb}
 \usepackage{xcolor}
\usepackage{bm} 
\expandafter\let\csname equation*\endcsname\relax

\expandafter\let\csname endequation*\endcsname\relax

\usepackage{amsmath}
\usepackage{physics}
\usepackage{stmaryrd}

\newcommand{\Z}{\mathbb{Z}}

\newcommand{\Var}{\text{Var}}

\let\oldref\ref
\renewcommand{\ref}[1]{(\oldref{#1})}

\usepackage{tikz}
\usetikzlibrary{calc}
\usetikzlibrary{shapes}
\usetikzlibrary{calc,decorations.markings,decorations.pathmorphing}

\begin{document}

\title[Number of distinct sites visited by a resetting random walker]{Number of distinct sites visited by a resetting random walker}

\author{Marco Biroli$\,^\dagger\,^\star$, Francesco Mori$\,^\star$, Satya N. Majumdar$\,^\star$}
\address{$\dagger$ Physics Department, \'Ecole normale sup\'erieure, Rue d'Ulm, 75005 Paris, France}
\address{$\,^\star$ LPTMS, CNRS, Univ. Paris-Sud, Universit\'e Paris-Saclay, 91405 Orsay, France}

\begin{abstract}
We investigate the number $V_p(n)$ of distinct sites visited by an $n$-step resetting random walker on a $d$-dimensional hypercubic lattice with resetting probability $p$. In the case $p=0$, we recover the well-known result that the average number of distinct sites grows for large $n$ as $\langle V_0(n)\rangle\sim n^{d/2}$ for $d<2$ and as $\langle V_0(n)\rangle\sim n$ for $d>2$. For $p>0$, we show that $\langle V_p(n)\rangle$ grows extremely slowly as $\sim \left[\log(n)\right]^d$. We observe that the recurrence-transience transition at $d=2$ for standard random walks (without resetting) disappears in the presence of resetting. In the limit $p\to 0$, we compute the exact crossover scaling function between the two regimes. In the one-dimensional case, we derive analytically the full distribution of $V_p(n)$ in the limit of large $n$. Moreover, for a one-dimensional random walker, we introduce a new observable, which we call \emph{imbalance}, that measures how much the visited region is symmetric around the starting position. We analytically compute the full distribution of the imbalance both for $p=0$ and for $p>0$. Our theoretical results are verified by extensive numerical simulations.
\end{abstract}


\maketitle

\section{Introduction}
\label{sec:intro}

Resetting random walks (RWs) have recently emerged as an active research field in the context of stochastic processes. They appeared in the literature only a decade ago but managed to find applications ranging from computer science to ecology - for a recent review see \cite{EvansReview}. For instance, the motion of foraging animals can be modeled as a resetting RW \cite{Boyer, Giuggioli}. Indeed, animals tend to go back to some fixed location (e.g., to their nest) when searching for food. Other examples appear in the context of search algorithms in computer science, where stochastic resetting has proved beneficial to the search process \cite{Luby,Montanari,  Tong, Gelenbe}, and in biology, e.g, to describe catastrophes in population dynamics \cite{VAM10}.

Moreover, resetting has been studied for a wide range of stochastic processes. Examples include Brownian motion \cite{Evans,EM11, Whitehouse, Montero,EM14, MajumdarSabhapandit,Pal2, Pal, Reuveni, Bhat,  Pal3,EM18,G19,DRR20}, random accelleration process \cite{S20}, L\'evy flights \cite{Kusmierz, Kusmierz2} and active particles \cite{EvansMajumdar2018, Masoliver, Kumar}. The resetting dynamics induce a net probability flux towards the resetting location, forcing the system out of equilibrium. Hence, resetting RWs are of particular interest to study nonequilibrium steady states \cite{EvansMajumdar2013, Gupta, Durang,Gupta2,  Eule}. Some variations of resetting RWs have also been studied such as multiple resetting locations \cite{Fernanda}, non-Poissonian resetting \cite{Bressloff}, resetting to the maximal position \cite{MSS15}, or resetting to a previously visited location \cite{BS14,BEM17,FBG17}. Resetting has also been considered many-body systems \cite{Gupta,Durang,BKP19,Magoni20}, quantum systems \cite{MSM18,RTLG18}, and optimal control theory \cite{DM21}. Moreover, several theoretical results, e.g., the mean first-passage time to a fixed target, have been verified in experiments using silica microspheres in optical traps \cite{BBP20,TPS20,FBP21,MajumdarConvex}. For a continuous-time resetting Brownian motion numerous exact results are already known, such as the first passage probability \cite{Evans} or the distribution of the value and time of the maximum in the one-dimensional case \cite{Evans, Mori, Singh}. 

One of the simplest and most fundamental quantities characterizing the spatial spread of an RW on a discrete graph is the number $V(n)$ of distinct visited sites after $n$ steps. For instance, in the case of foraging animals, $V(n)$ corresponds to the size of the territory covered by the animal. Similarly, in the case of graph search algorithms, this quantity corresponds to the number of locations that have already been explored. The problem of computing the average value of $V(n)$ was first introduced by Dvoretzky and Erd\"os in 1951 \cite{Erdos} and was solved in a number of later papers \cite{V63,MW65,Jain1, Jain2} in the case of an RW without resetting. Since then, these results have been extended to a variety of different processes \cite{Mori, Dayan, Larralde4,HLT92,B94, Larralde1,Larralde3, Berkolaiko,YA00,AY00,Larralde2, MT12, Klinger,KBV21}. For instance, in the case of $N$ independent walkers, both the number of distinct sites \cite{Larralde4,Berkolaiko,YA00,Larralde2,KMS13} and the number of common sites \cite{MT12} have been studied. Related quantities, such as the cover time of finite intervals, have also been investigated \cite{CBV15,MSS16}. Moreover, the number of visited sites up to an exit time from a finite domain was studied in the case of resetting RW in one-dimension \cite{Klinger,KBV21}. Recently, the perimeter and area of the convex hull, which gives an approximate measure of the space explored by the process, have been studied in Ref.~\cite{MajumdarConvex} for a two-dimensional Brownian motion with resetting. Here, we focus instead on a more direct measure of the space explored by considering the number of visited sites.



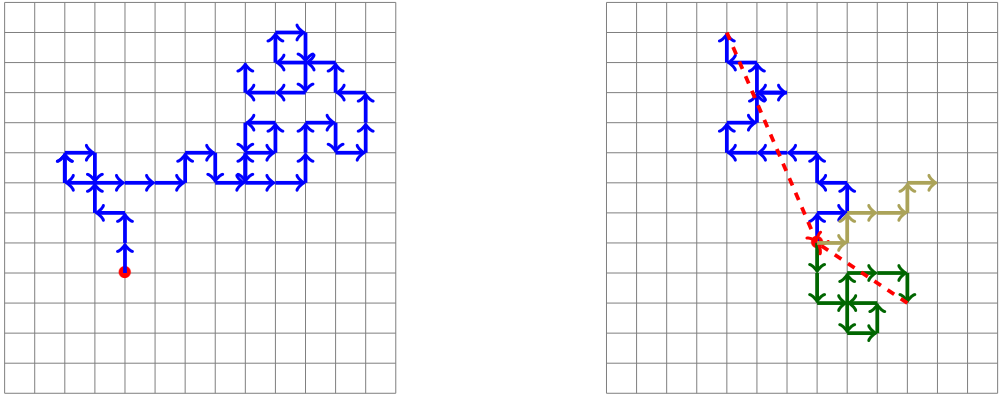
\begin{figure}
     \centering
     \begin{subfigure}[b]{0.49\textwidth}
         \centering
         \begin{tikzpicture}[scale=0.4]

    \foreach \i in {-4,...,9} {
        \draw [very thin,gray] (\i,-4) -- (\i,9);
    }
    \foreach \i in {-4,...,9} {
        \draw [very thin,gray] (-4,\i) -- (9,\i);
    }
\node [-,red] at (0,0) {\textbullet};

\draw [->,blue, line width=0.5mm]  (0,0)-- (0,1);
\draw [->,blue, line width=0.5mm]  (0,1)-- (0,2);
\draw [->,blue, line width=0.5mm]  (0,2)-- (-1,2);
\draw [->,blue, line width=0.5mm]  (-1,2)-- (-1,3);
\draw [->,blue, line width=0.5mm]  (-1,3)-- (-2,3);
\draw [->,blue, line width=0.5mm]  (-2,3)-- (-2,4);
\draw [->,blue, line width=0.5mm]  (-2,3)-- (-2,4);
\draw [->,blue, line width=0.5mm]  (-2,4)-- (-1,4);
\draw [->,blue, line width=0.5mm]  (-1,4)-- (-1,3);
\draw [->,blue, line width=0.5mm]  (-1,3)-- (0,3);
\draw [->,blue, line width=0.5mm]  (0,3)-- (1,3);
\draw [->,blue, line width=0.5mm]  (1,3)-- (2,3);
\draw [->,blue, line width=0.5mm]  (2,3)-- (2,4);
\draw [->,blue, line width=0.5mm]  (2,4)-- (3,4);
\draw [->,blue, line width=0.5mm]  (3,4)-- (3,3);
\draw [->,blue, line width=0.5mm]  (3,3)-- (4,3);
\draw [->,blue, line width=0.5mm]  (4,3)-- (4,4);
\draw [->,blue, line width=0.5mm]  (4,4)-- (5,4);
\draw [->,blue, line width=0.5mm]  (5,4)-- (5,5);
\draw [->,blue, line width=0.5mm]  (5,5)-- (4,5);
\draw [->,blue, line width=0.5mm]  (4,5)-- (4,4);
\draw [->,blue, line width=0.5mm]  (4,4)-- (4,3);
\draw [->,blue, line width=0.5mm]  (4,3)-- (5,3);
\draw [->,blue, line width=0.5mm]  (5,3)-- (6,3);
\draw [->,blue, line width=0.5mm]  (6,3)-- (6,4);
\draw [->,blue, line width=0.5mm]  (6,4)-- (6,5);
\draw [->,blue, line width=0.5mm]  (6,5)-- (7,5);
\draw [->,blue, line width=0.5mm]  (7,5)-- (7,4);
\draw [->,blue, line width=0.5mm]  (7,4)-- (8,4);
\draw [->,blue, line width=0.5mm]  (8,4)-- (8,5);
\draw [->,blue, line width=0.5mm]  (8,5)-- (8,6);
\draw [->,blue, line width=0.5mm]  (8,6)-- (7,6);
\draw [->,blue, line width=0.5mm]  (7,6)-- (7,7);
\draw [->,blue, line width=0.5mm]  (7,7)-- (6,7);
\draw [->,blue, line width=0.5mm]  (6,7)-- (5,7);
\draw [->,blue, line width=0.5mm]  (5,7)-- (5,8);
\draw [->,blue, line width=0.5mm]  (5,8)-- (6,8);
\draw [->,blue, line width=0.5mm]  (6,8)-- (6,7);
\draw [->,blue, line width=0.5mm]  (6,7)-- (6,6);
\draw [->,blue, line width=0.5mm]  (6,6)-- (5,6);
\draw [->,blue, line width=0.5mm]  (5,6)-- (4,6);
\draw [->,blue, line width=0.5mm]  (4,6)-- (4,7);

\end{tikzpicture}
        
     \end{subfigure}
     \hfill
     \begin{subfigure}[b]{0.49\textwidth}
         \centering
         \begin{tikzpicture}[scale=0.4]

    \foreach \i in {-7,...,6} {
        \draw [very thin,gray] (\i,-5) -- (\i,8);
    }
    \foreach \i in {-5,...,8} {
        \draw [very thin,gray] (-7,\i) -- (6,\i);
    }

\node [-,red] at (0,0) {\textbullet};

\draw [->,blue, line width=0.5mm]  (0,0)-- (0,1);
\draw [->,blue, line width=0.5mm]  (0,1)-- (1,1);
\draw [->,blue, line width=0.5mm]  (1,1)-- (1,2);
\draw [->,blue, line width=0.5mm]  (1,2)-- (0,2);
\draw [->,blue, line width=0.5mm]  (0,2)-- (0,3);
\draw [->,blue, line width=0.5mm]  (0,3)-- (-1,3);
\draw [->,blue, line width=0.5mm]  (-1,3)-- (-2,3);
\draw [->,blue, line width=0.5mm]  (-2,3)-- (-3,3);
\draw [->,blue, line width=0.5mm]  (-3,3)-- (-3,4);
\draw [->,blue, line width=0.5mm]  (-3,4)-- (-2,4);
\draw [->,blue, line width=0.5mm]  (-2,4)-- (-2,5);
\draw [->,blue, line width=0.5mm]  (-2,5)-- (-1,5);
\draw [->,blue, line width=0.5mm]  (-1,5)-- (-2,5);
\draw [->,blue, line width=0.5mm]  (-2,5)-- (-2,6);
\draw [->,blue, line width=0.5mm]  (-2,6)-- (-3,6);
\draw [->,blue, line width=0.5mm]  (-3,6)-- (-3,7);

\draw [->,red,dashed, line width=0.5mm]  (-3,7)-- (0,0);
\draw [->, black!60!green, line width=0.5mm]  (0,0)-- (0,-1);
\draw [->,black!60!green, line width=0.5mm]  (0,-1)-- (0,-2);
\draw [->,black!60!green, line width=0.5mm]  (0,-2)-- (1,-2);
\draw [->,black!60!green, line width=0.5mm]  (1,-2)-- (1,-3);
\draw [->,black!60!green, line width=0.5mm]  (1,-3)-- (2,-3);
\draw [->,black!60!green, line width=0.5mm]  (2,-3)-- (2,-2);
\draw [->,black!60!green, line width=0.5mm]  (2,-2)-- (1,-2);
\draw [->,black!60!green, line width=0.5mm]  (1,-2)-- (1,-1);
\draw [->,black!60!green, line width=0.5mm]  (1,-1)-- (2,-1);
\draw [->,black!60!green, line width=0.5mm]  (2,-1)-- (3,-1);
\draw [->,black!60!green, line width=0.5mm]  (3,-1)-- (3,-2);
\draw [->,red,dashed, line width=0.5mm]  (3,-2)-- (0,0);
\draw [->, black!40!yellow, line width=0.5mm]  (0,0)-- (1,0);
\draw [->, black!40!yellow, line width=0.5mm]  (1,0)-- (1,1);
\draw [->, black!40!yellow, line width=0.5mm]  (1,1)-- (2,1);
\draw [->, black!40!yellow, line width=0.5mm]  (2,1)-- (3,1);
\draw [->, black!40!yellow, line width=0.5mm]  (3,1)-- (3,2);
\draw [->, black!40!yellow, line width=0.5mm]  (3,2)-- (4,2);

\end{tikzpicture}
         
     \end{subfigure}
        \caption{Typical trajectories for non-resetting (left panel) and resetting (right panel) random walks on a two-dimensional square lattice. In the right panel, different colors correspond to trajectories between different resetting events, which are indicated by red dashed lines.}
        \label{fig:ex_walks}
\end{figure}

In this paper, we investigate the number $V_p(n)$ of distinct visited sites by a $n$-step RW with resetting probability $p$ moving on a $d$-dimensional hypercubic lattice, whose sites are identified with the integers $\mathbb{Z}^d$. Note that we have introduced the subscript $p$ in $V_p(n)$ to stress the dependence on the resetting probability. Let $\vb{X}(n)$ denote the position of the walker after $n$ steps, starting from the origin $\vb{X}(0)=\vb{0}$. The position of the walker at step $n$ evolves according to
\begin{equation}\label{eq:def_X}
\vb{X}(n+1) = 
\begin{cases}
\vb{0}\,, &\mbox{~~with probability~~} p\,,\\
\vb{X}(n) \pm  \vb{e_i} \,, &\mbox{~~with probabiliy~~} \frac{1}{2d}(1 - p)\,,\\
\end{cases}\,,
\end{equation}
where $\vb{e_i}$, with $i = 1, \cdots, d$, are $d$-dimensional orthogonal unit vectors and $0\leq p\leq1$ is the resetting probability. At each time step, with probability $1-p$ the random walker jumps to a randomly chosen neighboring site. With the complementary probability $p$, the RW is reset to its starting position $\vb{0}$. For $p=0$ the walker is never reset and one recovers the usual RW model. Typical trajectories of RWs with and without resetting are shown in Fig. \ref{fig:ex_walks}. The main goal of this paper is to investigate the behavior of $V_p(n)$ in the limit of large $n$.

The rest of this paper is organized as follows. In Section \ref{sec:results} we provide a short summary of our main results. In Section \ref{sec:avg} we investigate the average value of the number of different sites in arbitrary dimension. Specifically, in Section \ref{sec:p0} we present a derivation of the classical results valid for RWs without resetting, in Section \ref{sec:p1} we focus instead on the case with resetting, and in Section \ref{sec:scaling} we investigate the scaling limit which interpolates between the two regimes. In Section \ref{sec:one-dim}, considering a continuous-time setting, we derive the full distribution of the number of visited sites in one dimension. Finally, in Section \ref{sec:conclusion} we conclude with final remarks and future direction. Some details of the computations are presented in the appendices.

\subsection{Summary of the main results}
\label{sec:results}

In this section, we summarize the main results of the paper. It is instructive to first consider the case of an RW without resetting, corresponding to $p=0$. In this case, we reproduce the well-known result \cite{Dayan,Larralde4,HLT92}
\begin{equation}
\langle V_0(n)\rangle\sim \begin{cases}
n^{d/2}\,\,\,&\text{for  }d<2\\
n/\log(n)\,\,\,&\text{for  }d=2\\
n\,\,\,&\text{for  }d>2\,.\\
\end{cases}
\label{eq:V0n_intro}
\end{equation}
This result is a direct consequence of the recurrence-transience transition of RWs \cite{Book}. For $d\leq 2$ the RW is recurrent, meaning that it visits its initial location infinitely often for $n\to \infty$. In this case, visiting a new site becomes hard at late times and the average number of distinct visited sites grows as $n^{d/2}$, with logarithmic corrections for $d=2$. Conversely, for $d>2$ the RW is said to be transient, meaning that each site is visited only a finite number of times for $n\to \infty$. As a consequence, the number of distinct sites grows linearly in $n$, meaning that almost every site that the RW visits is actually visited for the first time. Note that $\langle V_0(n)\rangle$ cannot grow faster than linearly in $n$. As we will show, this transition disappears when the resetting process is turned on.

In the case of resetting RWs, corresponding to $p>0$, we show that for $n\gg1$
\begin{equation}
\langle V_p(n)\rangle \approx \mathcal{A}(d)\int_{0}^{\infty} dR~R^{d-1}\left[1-\exp\left(-\frac{n}{C_d(p) R^{(d-1)/2}\exp\left(\sqrt{\frac{2dp}{1-p}}R\right)}\right)\right]\,,
\label{Vp_full_intro}
\end{equation}
where 
\begin{equation}\label{eq:def_Ad}
\mathcal{A}(d) = \frac{2 \pi^{d/2}}{\Gamma(d/2)}\,,
\end{equation}
$\Gamma(z)$ is the gamma function. The constant $C_d(p)$, given in Eq.~\ref{C_dp}, depends on the details of the lattice and can be computed by numerical integration. This asymptotic result in Eq.~\ref{Vp_full_intro} is shown in Fig.~\ref{fig:V} for $d=1,2,3$, and is in perfect agreement with numerical simulations. For large $n$, the leading order behavior of the expression in Eq.~\ref{Vp_full_intro} is given by
\begin{equation}
\langle V_p(n)\rangle\approx\frac{\mathcal{A}(d)}{d}\left(\frac{1-p}{2pd}\right)^{d/2}\left[\log(n)\right]^d+\mathcal{O}\left(\left[\log(n)\right]^{d-1}\right)\,,
\label{eq:Vpn_intro}
\end{equation}
Interestingly, the recurrence-transience transition disappears once we switch on the resetting probability and we obtain the unique asymptotic expression in Eq.~\ref{eq:Vpn_intro} valid for any $d>0$. For late times, the resetting random walker will reach a steady state distribution, which is centered around the resetting location. For this reason, visiting new sites far away from the resetting location becomes exponentially rare in time, leading to the slow logarithmic growth of $\langle V_p(n)\rangle$. Note that the leading order behavior in Eq.~\ref{eq:Vpn_intro} is completely independent of $C_d(p)$, suggesting that this result is independent of the details of the lattice and valid for any regular lattice in $d$ dimension.

Note that the expression in Eq.~\ref{eq:Vpn_intro} breaks down for $p\to 0$. For $d<2$, the crossover between the two regimes in Eqs.~\ref{eq:V0n_intro} and \ref{eq:Vpn_intro} is described by the scaling limit $n\to \infty$, $p\to 0$ with $np$ fixed. In this limit, we show
\begin{equation}
\langle V_p(n)\rangle\approx \frac{\mathcal{A}(d)}{(2dp)^{d/2}}F_d(pn)\,,
\label{scaling_intro}
\end{equation}
where $F_d(z)$ is a dimension-dependent scaling function that we compute explicitly for any $d$. On the other hand, for $d>2$ the crossover occurs in the limit $p\to 0$ and $n\to \infty$ with $np^{d/2}$ fixed, where
\begin{equation}
\langle V_p(n)\rangle\approx \frac{\mathcal{A}(d)}{(2dp)^{d/2}}F_d(np^{d/2})\,,
\label{scaling_intro_d>2}
\end{equation}
In the limit case $d=2$, we find that the correct scaling regime is $n\to\infty$, $p\to 0$, with $pn/\log(1/p)$ fixed, with the scaling form
\begin{equation}
\langle V_p(n)\rangle\approx \frac{\pi}{2p}F_2\left(\frac{pn}{\log(1/p)}\right)\,.
\label{scaling_intro_d2}
\end{equation}
The exact expression of $F_d(z)$ for $d=1$, $d=2$, and $d=3$ are respectively given in Eqs.~\ref{eq:f1z}, \ref{f2}, and \ref{eq:f3z}. The scaling forms in Eqs.~\ref{scaling_intro}, \ref{scaling_intro_d>2}, and \ref{scaling_intro_d2} interpolate between the behaviors in Eqs.~\ref{eq:V0n_intro} and \ref{eq:Vpn_intro}. Indeed, the scaling function $F_d(z)$ behaves for small $z$ as
\begin{equation}
F_d(z)\sim\begin{cases}
z^{d/2}\,,\quad &\text{ for }d<2\,,\\
\\
z\,,\quad &\text{ for }d\geq 2\,,\\
\end{cases}
\end{equation}
in agreement with the case without resetting in Eq.~\ref{eq:V0n_intro}. On the other hand, for large $z$, we find
\begin{equation}
F_d(z)\approx\frac{1}{d}\left[\log(z)\right]^d\,,
\end{equation}
for any $d>0$, reproducing the behavior in Eq.~\ref{eq:Vpn_intro}. In other words, the slow logarithmic growth appears after $n^*$ steps, where $n^*\sim p^{-1}$ for $d<2$, $n^*\sim \log(1/p)/p$ for $d=2$, and $n^*\sim p^{-d/2}$ for $d>2$. These exact results for $F_1(z)$, $F_2(z)$, and $F_3(z)$ are shown in Fig.~\ref{fig:V_scaling} and are in good agreement with numerical simulations.

Notably, in the one-dimensional case, one can compute the full distribution of the number of distinct sites. Indeed, for a one-dimensional random walker, the number of visited sites coincides with the \emph{span} of the process, defined as the difference between the global maximum and the global minimum. Note that this quantity, which we precisely define below, is also sometimes called \emph{range} in the extreme value theory literature. For simplicity, we focus on the continuous-time limit where the resetting RW can be approximated as a resetting Brownian motion (BM). To consider this limit, we introduce the length $a>0$ and the duration $\Delta t>0$ of a step. The continuum limit corresponds to the limit $a,\Delta t,p\to 0$ and $n\to \infty$, with $r=p/ \Delta t$, $D=a^2/(2\Delta t)$, and $t=n\Delta t$ fixed. The constant $r$ is the resetting rate of the process, $D$ is the diffusion coefficient, and $t$ is the total time. We denote by $x(t)=a X(t/\Delta t)$ the position of the process at time $t$, with $x(0)=0$.

The span of the process at time $t$ is defined as
\begin{equation}
L(t)=M(t)-m(t)\,,
\end{equation}
where $M(t)=\max_{\tau<t}x(\tau)$ is the maximum of the process and $m(t)=\min_{\tau<t}x(\tau)$ is the minimum. Note that by definition $m(t)<0$. Then, it is easy to show that the number of visited sites is related to the span by
\begin{equation}
V_p(n)\to\frac{1}{a}L(t=n\Delta t)\,.
\end{equation}
Note that the span $L(t)$ has the dimension of a length while $V_p(n)$ is dimensionless. The distribution of the span is a central quantity in the context of extreme value theory \cite{EVS_review} and has been investigated for a wide range of stochastic processes \cite{Klinger,KBV21,KMS13,F51,RMS15,W20,GSM21}. Here, we show that for $rt\gg 1$, the distribution $P_r(L,t)$ of the span $L(t)$ of a BM with resetting rate $r$ assumes the scaling form
\begin{equation}
P_r(L, t) \approx  \sqrt{r/D} ~g\left[\sqrt{r/D}(L-2\sqrt{D/r}\log(rt))\right]\,,
\label{PL_scale_intr}
\end{equation}
where 
\begin{equation}
g(z) = 2e^{-z} K_0\left(2e^{-z/2}\right)\,.
\label{g_intro}
\end{equation}
Here $K_0(z)$ is the modified Bessel function of the second kind. This scaling function has asymptotic behaviors
\begin{equation}
g(z)\approx\begin{cases}
\sqrt{\pi}\exp\left[-2e^{-z/2}-\frac34 z\right]~~~&\text{for }z\to -\infty\\
\\
ze^{-z}~~~&\text{for }z\to \infty\,.\\
\end{cases}
\label{g_asympt}
\end{equation}
Since $g(z)$ is positive and normalized to unity in $-\infty<z<\infty$, this result can be interpreted as follows. For $t\gg 1/r$ the span can be written as
\begin{equation}
L(t)\approx 2\sqrt{\frac{D}{r}}\log(rt)+\sqrt{\frac{D}{r}}z\,,
\end{equation}
where $z$ is an random variable with probability density function (PDF) $g(z)$. In Section \ref{sec:one-dim} we provide an intuitive interpretation of this result based on extreme value theory. From this distribution we find that the average number of distinct sites goes for $rt \gg 1$ as
\begin{equation} \label{eq:3rd-result}
\langle L(t) \rangle \approx 2\sqrt{\frac{D}{r}} (\log(rt) + \gamma_E),
\end{equation}
in agreement with \ref{eq:Vpn_intro} for $d=1$. Here $\gamma_E=0.57721\ldots$ is the Euler constant. Similarly, the variance is given by
\begin{equation}\label{eq:4th-result}
\Var(L(t)) \approx \frac{ \pi^2}{3}\frac{D }{r}.
\end{equation}

Finally, having investigated the size of the region visited by the walker, one might ask what is the typical shape of this region. In particular, in one dimension, it is relevant to investigate whether or not the walker visits the positive and the negative $x$-axes symmetrically. To address this question, we introduce the \emph{imbalance} $l(t)$, which we define as
\begin{equation}
l(t)=M(t)+m(t)\,,
\end{equation}
where we recall that $M(t)>0$ and $m(t)<0$ are respectively the maximum and the minimum of the process up to time $t$. Interestingly, $l(t)$ is positive (negative) when the walker has mainly visited the sites to the right (left) of its starting position $x(0)=0$. To the best of our knowledge, this quantity $l(t)$ has never been analytically investigated.

In the case of BM without resetting and with diffusion coefficient $D$, we show that the PDF $P_0(l,t)$ of the imbalance $l(t)$ is given for any $t$ by
\begin{equation}
P_0(l,t)=\frac{1}{\sqrt{Dt}}\mathcal{F}\left(\frac{l}{\sqrt{Dt}}\right)\,,
\end{equation}
where the scaling function $\mathcal{F}(z)$ is rather nontrivial even for a one-dimensional BM and is given in Eq.~\ref{F}. As a consequence of the $x\to-x$ symmetry of BM, $P_0(l,t)$ is symmetric around $l=0$ and the first moment vanishes $\langle l(t)\rangle=0$. The second moment reads
\begin{equation}
\langle l(t)^2\rangle=8\log(\frac{e}{2})Dt\,.
\end{equation}
Moreover, the scaling function $\mathcal{F}(z)$ has asymptotic behaviors \ref{app:fl}
\begin{equation}
\mathcal{F}(z)\approx\begin{cases}
\sqrt{\pi}/8-\sqrt{\pi}z^2/128 \,,&\text{ for }|z|\to 0\,,\\
\\
\frac{2}{3\sqrt{\pi}}e^{-z^2/4}\,,&\text{ for }|z|\to \infty\,.
\end{cases}
\label{F_z_asymp}
\end{equation}
The distribution of $l(t)$ is shown in Fig.~\ref{fig:P0l} and is in good agreement with numerical simulation. We observe that the PDF of $l$ is maximal for $l=0$ and it decreases exponentially for large $l$.

In the case where resetting is present, we show that for $rt\gg 1$ 
\begin{equation}
P_r(l,t)\approx \sqrt{\frac{r}{D}}h\left(\sqrt{\frac{r}{D}}\,\,l\right)\,,
\label{Pr_lt_intro}
\end{equation}
where
\begin{equation}
h(y)=\frac{1}{4\cosh^2(y/2)}\,.
\label{h_intro}
\end{equation}
This scaling function $h(y)$ has asymptotic behaviors
\begin{equation}
h(y)\approx
\begin{cases}
1/4-y^2/16\,,\quad\text{  for  }|y|\to 0\,,\\
\\
e^{-y}\,,\quad\text{  for  }|y|\to \infty\,.\\
\end{cases}
\label{h_asymp}
\end{equation}
The analytical prediction for $P_r(l,t)$ for large $t$, given in Eqs.~\ref{Pr_lt_intro} and \ref{h_intro}, is shown in Fig.~\ref{fig:1d_l_pdf} and is in good agreement with numerical simulations. We also derive the mean and variance of $l$ in the long-time limit which are given by
\begin{equation} 
\langle l(t) \rangle = 0 \mbox{~~and~~} \Var(l(t)) = \frac{\pi^2}{3}\frac{D}{r}\,.
\end{equation}
The result in Eq.~\ref{Pr_lt_intro} indicates that for late times the imbalance distribution does not depend on time. This shows that, while the size of the visited region grows as $\log(rt)$, the RW visits symmetrically (up to an order-one correction) the sites to the left and to the right of the origin. In other words, the resetting process makes the set of visited sites symmetric around the resetting location.

\section{Average number of distinct sites}

\label{sec:avg}

In this section, we investigate the average number $\langle V_p(n)\rangle$ of distinct sites visited by an $n$-step RW with resetting probability $p$. The starting point of our derivation is the identity
\begin{equation}
V_p(n)=\sum_{\vb{X}}\sigma(\vb{X},n)\,,
\label{identity_sigma}
\end{equation}
where the sum over $\vb{X}$ runs over all elements of $\Z^d$ and $\sigma(\vb{X},n)$ are binary variables defined as
\begin{equation}
\sigma(\vb{X},n)=\begin{cases}
1\,,&\text{       if $\vb{X}$ is visited before step $n$,}\\
0\,,&\text{       otherwise.}
\end{cases}
\label{definition_sigma}
\end{equation}
Considering the average value on both sides of Eq.~\ref{identity_sigma}, we obtain
\begin{equation}
\langle V_p(n)\rangle=\sum_{\vb{X}}\text{Prob.(site $\vb{X}$ is visited before step $n$)}\,.
\label{identity_sigma2}
\end{equation}
In turn, the probability that a given site is visited at least once in $n$ steps can be expressed by partitioning on the time $m$ of the first visit as
\begin{equation}
\text{Prob.(site $\vb{X}$ is visited before step $n$)}=\sum_{m=1}^n F_p(\vb{0},\vb{X},m)\,,
\label{identity_sigma3}
\end{equation}
where $F_p(\vb{0},\vb{X},m)$ is the probability that the site $\vb{X}$ is visited for the first time at step $m$, having started from the origin $\vb{0}$. The quantity $F_p(\vb{0},\vb{X},m)$ is usually referred to as \emph{first-passage probability}. Combining the expressions in Eqs.~\ref{identity_sigma2} and \ref{identity_sigma3}, we find
\begin{equation}
\langle V_p(n)\rangle=\sum_{\vb{X}}\sum_{m=1}^n F_p(\vb{0},\vb{X},m)\,.
\label{identity_sigma4}
\end{equation}
Notably, the first-passage probability $F_p(\vb{0},\vb{X},m)$ can be written in terms of a simpler quantity, i.e., the propagator $G_p(\vb{X}_0,\vb{X},n)$, defined as the probability to reach position $\vb{X}$ in $n$ steps starting from $\vb{X}_0$ for an RW with resetting probability $p$. Indeed, the propagator $G_p(\vb{X}_0,\vb{X},n)$ satisfies the relation, for $\vb{X}\neq \vb{X}_0$,
\begin{equation}
G_p(\vb{X}_0,\vb{X},n)=\sum_{m=1}^{n}F_p(\vb{X}_0,\vb{X},m)G_p(\vb{X},\vb{X},n-m)\,,
\label{relation_G_F}
\end{equation}
meaning that, in order to reach the site $\vb{X}$ at step $n$, the walker has to reach $\vb{X}$ for the first time at some intermediate step $m$ and then to come back to $\vb{X}$ after $n-m$ steps. Taking the generating function with respect to $n$ on both sides of Eq.~\ref{relation_G_F}, we obtain
\begin{equation}
\tilde{F}_p(\vb{X}_0,\vb{X},z)=\frac{\tilde{G}_p(\vb{X}_0,\vb{X},z)}{\tilde{G}_p(\vb{X},\vb{X},z)}\,,
\label{relation_G_F_2}
\end{equation}
where we have defined the generating functions
\begin{equation}
\tilde{G}_p(\vb{X}_0,\vb{X},z)=\sum_{n=0}^{\infty}G_p(\vb{X}_0,\vb{X},n)z^n\,,
\end{equation}
and 
\begin{equation}
\tilde{F}_p(\vb{X}_0,\vb{X},z)=\sum_{m=1}^{\infty}F_p(\vb{X}_0,\vb{X},m)z^m\,.
\end{equation}
On the other hand, in the case $\vb{X}=\vb{X}_0$ the relation in Eq.~\ref{relation_G_F} becomes
\begin{equation}
G_0(\vb{X}_0,\vb{X}_0,n)=\delta_{n,0}+\sum_{m=1}^{\infty}F_0(\vb{X}_0,\vb{X}_0,m)G_0(\vb{X}_0,\vb{X}_0,n)\,.
\end{equation}
Passing to the generating functions, we obtain
\begin{equation}
\tilde{F}_0(\vb{X}_0,\vb{X}_0,z)=1-\frac{1}{\tilde{G}_0(\vb{X}_0,\vb{X}_0,z)}\,.
\label{Fx0x0}
\end{equation}

Finally, taking a generating function with respect to $n$ on both sides of Eq.~\ref{identity_sigma4} and using the expression for $\tilde{F}_p(\vb{X}_0,\vb{X},z)$, given in Eq.~\ref{relation_G_F_2}, we obtain
\begin{equation}
\langle \tilde{V}_p(z)\rangle=\frac{1}{1-z}\sum_{\vb{X}}\frac{\tilde{G}_p(\vb{0},\vb{X},z)}{\tilde{G}_p(\vb{X},\vb{X},z)}\,,
\label{identity_V}
\end{equation}
where
\begin{equation}
\tilde{V}_p(z)=\sum_{n=1}^{\infty}V_p(n)z^n\,.
\end{equation}
Eq.~\ref{identity_V} is the first main result of this paper and relates the average number of visited sites to the propagator of the process. This relation turns out to be very general and the rest of this section is devoted to the analysis of this result in different cases. We will first consider the simpler case of RW without resetting, corresponding to $p=0$. Then, we will investigate the role of resetting in the case $p>0$.

\subsection{Random walks without resetting ($p=0$)}

\label{sec:p0}

In this subsection, we first reproduce the well-known results \cite{Dayan,Larralde4,HLT92} for $\langle V_0(n)\rangle$ for the RW without resetting ($p=0$ case). This will allow the reader to familiarize with the techniques to extract asymptotic large-$n$ behaviors from generating functions and also to appreciate our new results derived in Section \ref{sec:p1} for $p>0$, which are drastically different from the $p=0$ case.

 In the case $p=0$, the propagator $G_0(\vb{0},\vb{X},n)$ satisfies the recursion relation
\begin{equation}
G_0(\vb{0},\vb{X},n)=\frac{1}{2d}\sum_{i=1}^{d}\left[G_0(\vb{0},\vb{X}+\vb{e}_i,n-1)+G_0(\vb{0},\vb{X}-\vb{e}_i,n-1)\right]\,,
\label{eq:recursion}
\end{equation}
where $\vb{e}_i$ is the unit vector in the $i$-th direction and $1/(2d)$ is the probability of each jump. Note that in the case $p=0$ the system is translational invariant and one has $G_0(\vb{X}_0,\vb{X},n)=G_0(\vb{0},\vb{X}_0-\vb{X},n)$.  Solving this recursion \ref{app:rec}, we show that
\begin{equation}
G_0(\vb{0},\vb{X},n)=\int_{-\pi}^{\pi}\ldots\int_{-\pi}^{\pi}\frac{d^d\vb{k}}{(2\pi)^d}e^{-i \vb{k}\cdot\vb{X}}\left[\frac{1}{d}\sum_{i=1}^{d}\cos k_i\right]^n\,.
\label{eq:G0_expression}
\end{equation}
Considering the generating function with respect to $n$, we find
\begin{equation}
\tilde{G}_0(\vb{0},\vb{X},z)=\int_{-\pi}^{\pi}\ldots\int_{-\pi}^{\pi}\frac{d^d\vb{k}}{(2\pi)^d}e^{-i \vb{k}\cdot\vb{X}}\frac{1}{1-(z/d)\sum_{i=1}^{d}\cos k_i}\,.
\label{eq:Gz0_expression}
\end{equation}

Setting $p=0$ in the relation in Eq.~\ref{identity_V}, we obtain
\begin{equation}
\langle \tilde{V}_0(z)\rangle=\frac{1}{1-z}\sum_{\vb{X}}\frac{\tilde{G}_0(\vb{0},\vb{X},z)}{\tilde{G}_0(\vb{X},\vb{X},z)}\,.
\end{equation}
Using the translational invariance property we have $\tilde{G}_0(\vb{X},\vb{X},z)=\tilde{G}_0(\vb{0},\vb{0},z)$ and therefore
\begin{equation}
\langle \tilde{V}_0(z)\rangle=\frac{1}{(1-z)\tilde{G}_0(\vb{0},\vb{0},z)}\sum_{\vb{X}}\tilde{G}_0(\vb{0},\vb{X},z)\,.
\label{rel_V_G0}
\end{equation}
The sum over $\vb{X}$ can be now easily computed since
\begin{equation}
\sum_{\vb{X}}\tilde{G}_0(\vb{0},\vb{X},z)=\sum_{n=0}^{\infty}z^n~\sum_{\vb{X}}G_0(\vb{0},\vb{X},n)=\sum_{n=0}^{\infty}z^n=\frac{1}{1-z}\,,
\end{equation}
where we have used the fact that $G_0(\vb{0},\vb{X},n)$ is normalized to unity. Therefore, Eq.~\ref{rel_V_G0} can be rewritten as
\begin{equation}
\langle \tilde{V}_0(z)\rangle=\frac{1}{(1-z)^2\tilde{G}_0(\vb{0},\vb{0},z)}\,.
\label{rel_V_G0_1}
\end{equation}

We are interested in extracting the large-$n$ asymptotic behavior of $V_0(n)$ from this expression. To investigate this limit, we substitute $z=1-s$ and expand for small $s$. In this limit, the generating function on left-hand side of Eq.~\ref{rel_V_G0_1} can be approximated as
\begin{equation}
\langle \tilde{V}_0(z=1-s)\rangle=\sum_{n=0}^{\infty}\langle V_0(n)\rangle(1-s)^n\approx \int_{0}^{\infty}dn~\langle V_0(n)\rangle e^{-sn}\,.
\end{equation}
Considering the number $n$ of steps as a continuous variable, the generating function becomes a Laplace transform with Laplace variable $s$. Applying this approximation to Eq.~\ref{rel_V_G0_1}, we obtain
\begin{equation}
\int_{0}^{\infty}dn~\langle V_0(n)\rangle e^{-sn}\approx\frac{1}{s^2\tilde{G}_0(\vb{0},\vb{0},1-s)}\,.
\label{rel_V_G1}
\end{equation}
Thus, the late-time behavior of $\langle V_0(n)\rangle$ is determined by the small-$s$ expansions of $\tilde{G}_0(\vb{0},\vb{0},1-s)$. Using the exact expression in Eq.~\ref{eq:Gz0_expression}, we find that for small $s$ \ref{app:G_lim}
\begin{equation}
\tilde{G}_0(\vb{0},\vb{0},1-s)\approx\begin{cases}
A_d/s^{(2-d)/2}\,,&\text{ for }d<2\,,\\ 
\log(1/s)/\pi\,,&\text{ for }d=2\,,\\ 
B_d\,,&\text{ for }d>2\,,\\ 
\end{cases}
\label{G0_asymp}
\end{equation} 
where 
\begin{equation}
A_d=\frac{2^{-d/2}\pi^{1-d/2}d^{d/2}\csc(\frac{\pi d}{2})}{\Gamma(\frac{d}{2})}\,,
\label{A_d}
\end{equation}
and
\begin{equation}
B_d=\tilde{G}_0(\vb{0},\vb{0},1)\,.
\label{B_d}
\end{equation}
Note that $B_d<\infty$ for $d>2$. For instance, by numerical integration we find $B_3=1.516\ldots$ in the case $d=3$. 

Plugging the asymptotic behavior of $\tilde{G}_0(\vb{0},\vb{0},1-s)$ for $d<2$ into Eq.~\ref{rel_V_G1}, we obtain
\begin{equation}
\int_{0}^{\infty}dn~\langle V_0(n)\rangle e^{-sn}\approx\frac{1}{A_d s^{d/2+1}}\,.
\end{equation}
Inverting the Laplace transform, we find that for $d<2$ and $n\gg 1$
\begin{equation}
\langle V_0(n)\rangle \approx\frac{1}{A_d\Gamma(1+d/2)}n^{d/2}\,.
\label{V0_d<2}
\end{equation}
In particular, for a one-dimensional RW we obtain
\begin{equation}
\langle V_0(n)\rangle\approx 2\sqrt{\frac{2 n}{\pi}}\,.
\label{V0_d1}
\end{equation}
We have recovered the classical result that the number of visited sites in one dimension grows as the square root of the number of steps. Conversely, for $d>2$ we obtain
\begin{equation}
\int_{0}^{\infty}dn~\langle V_0(n)\rangle e^{-sn}\approx\frac{1}{B_d s^2}\,.
\end{equation}
Inverting the Laplace transform, we find that for large $n$
\begin{equation}
\langle V_0(n)\rangle\approx \frac{n}{B_d}\,,
\label{V0_d>2}
\end{equation}
meaning that the number of visited sites grows linearly with the number of steps for $d>2$. Finally, in the limiting case $d=2$ we have
\begin{equation}
\int_{0}^{\infty}dn~\langle V_0(n)\rangle e^{-sn}\approx\frac{\pi}{ s^2\log(1/s)}\,.
\label{d2_LT}
\end{equation}
Inverting the Laplace transform \ref{app:laplace_inv}, we obtain
\begin{equation}
\langle V_0(n)\rangle \approx\pi \frac{n}{\log(n)}\,,
\label{V0_d=2}
\end{equation}
corresponding to a linear growth with logarithmic corrections. The different behaviors of $\langle V_0(n)\rangle$ are summarized in Eq.~\ref{eq:V0n_intro} and are a consequence of the recurrence-transience transition \cite{Book}.

To understand the mechanism of this transition, it is useful to compute the probability $P_d$ that an RW eventually, i.e., with arbitrary many steps, returns to its starting position. Denoting the starting position by $\vb{X}_0$, the return probability $P_d$ can be expressed in terms of the first-passage probability as
\begin{equation}
P_d=\sum_{m=1}^{\infty}F_0(\vb{X}_0,\vb{X}_0,m)\,.
\end{equation}
We recall the result in Eq.~\ref{Fx0x0}, which reads
\begin{equation}
\tilde{F}_0(\vb{X}_0,\vb{X}_0,z)=\sum_{m=1}^{\infty}F_0(\vb{X}_0,\vb{X}_0,m)z^m=1-\frac{1}{\tilde{G}_0(\vb{X}_0,\vb{X}_0,z)}\,.
\end{equation}
Taking the limit $z\to 1$ and using the definition of $P_d$ above, we obtain
\begin{equation}
P_d=\lim_{z\to1}\left[1-\frac{1}{\tilde{G}_0(\vb{X}_0,\vb{X}_0,z)}\right]\,.
\end{equation}
Interestingly, for $d\leq 2$, the propagator $\tilde{G}_0(\vb{X}_0,\vb{X}_0,z)$ diverges for $z\to1$ (see Eq.~\ref{G0_asymp}) and hence 
\begin{equation}
P_d=1\,,
\end{equation}
meaning that the RW will eventually return to its starting position. For this reason, RWs in $d\leq 2$ are said to be recurrent. Actually, it is easy to show that the RW will return to its starting position infinitely many times. On the other hand, for $d>2$, $\tilde{G}_0(\vb{X}_0,\vb{X}_0,1)$ is finite, yielding
\begin{equation}
P_d=1-\frac{1}{\tilde{G}_0(\vb{X}_0,\vb{X}_0,1)}<1\,.
\end{equation}
In other words, with finite probability the walker never returns to its starting position. For this reason, RWs in $d>2$ are said to be transient. As a consequence of this property, RWs in $d\leq 2$ explore a smaller region of space and the number of visited sites grows slower than linearly.

\subsection{Random walks with resetting ($p>0$)}
\label{sec:p1}

In this section, we derive the main results of this paper, valid for random walks with stochastic resetting. In particular, we show that the recurrence-transience transition disappears once the resetting probability is switched on. In order to exploit the relation in Eq.~\ref{identity_V}, we need an explicit expression for the propagator $G_p(\vb{X}_0,\vb{X},n)$ in the presence of resetting. Luckily, this quantity can be related to the propagator $G_0(\vb{X}_0,\vb{X},n)$ of RWs without resetting. Indeed, these propagators satisfy the renewal equation
\begin{equation}
G_p(\vb{X}_0,\vb{X},n)=(1-p)^n G_0(\vb{X}_0,\vb{X},n)+p\sum_{m=0}^{n}(1-p)^m G_0(\vb{0},\vb{X},m)\,.\label{G0_GP}
\end{equation}
The first term on the right-hand side corresponds to the case where no resetting occurs. In such a case, which happens with probability $(1-p)^n$, the propagator reduces to the one of the RW without resetting. The second term instead corresponds to the case where at least one resetting occurs. The index $m=n-n_{\rm last}$, where $n_{\rm last}$ denotes the step at which the last resetting occurs before step $n$. After the last resetting, the walker has to reach the site $\bm{X}$ starting from the origin in the remaining $m$ steps. Passing to the generating functions, the relation in Eq.~\ref{G0_GP} reduces to
\begin{equation}
\tilde{G}_p(\vb{X}_0,\vb{X},z)=\tilde{G}_0(\vb{X}_0,\vb{X},z(1-p))+\frac{p}{1-z} \tilde{G}_0(\vb{0},\vb{X},z(1-p))\,.
\end{equation}
Plugging this expression for $\tilde{G}_p(\vb{X}_0,\vb{X},z)$ into Eq.~\ref{identity_V}, we obtain
\begin{equation}
\langle \tilde{V}_p(z)\rangle=\frac{1-z+p}{1-z}\sum_{\vb{X}}\left[p+(1-z)\frac{\tilde{G}_0(\vb{0},\vb{0},z(1-p))}{\tilde{G}_0(\vb{0},\vb{X},z(1-p))}\right]^{-1}\,,
\label{identity_Vp}
\end{equation}
As a check, it is easy to verify that by setting $p=0$ one recovers the expression in Eq.~\ref{rel_V_G0}.

\begin{figure}
\centering
\includegraphics[width=0.5\textwidth]{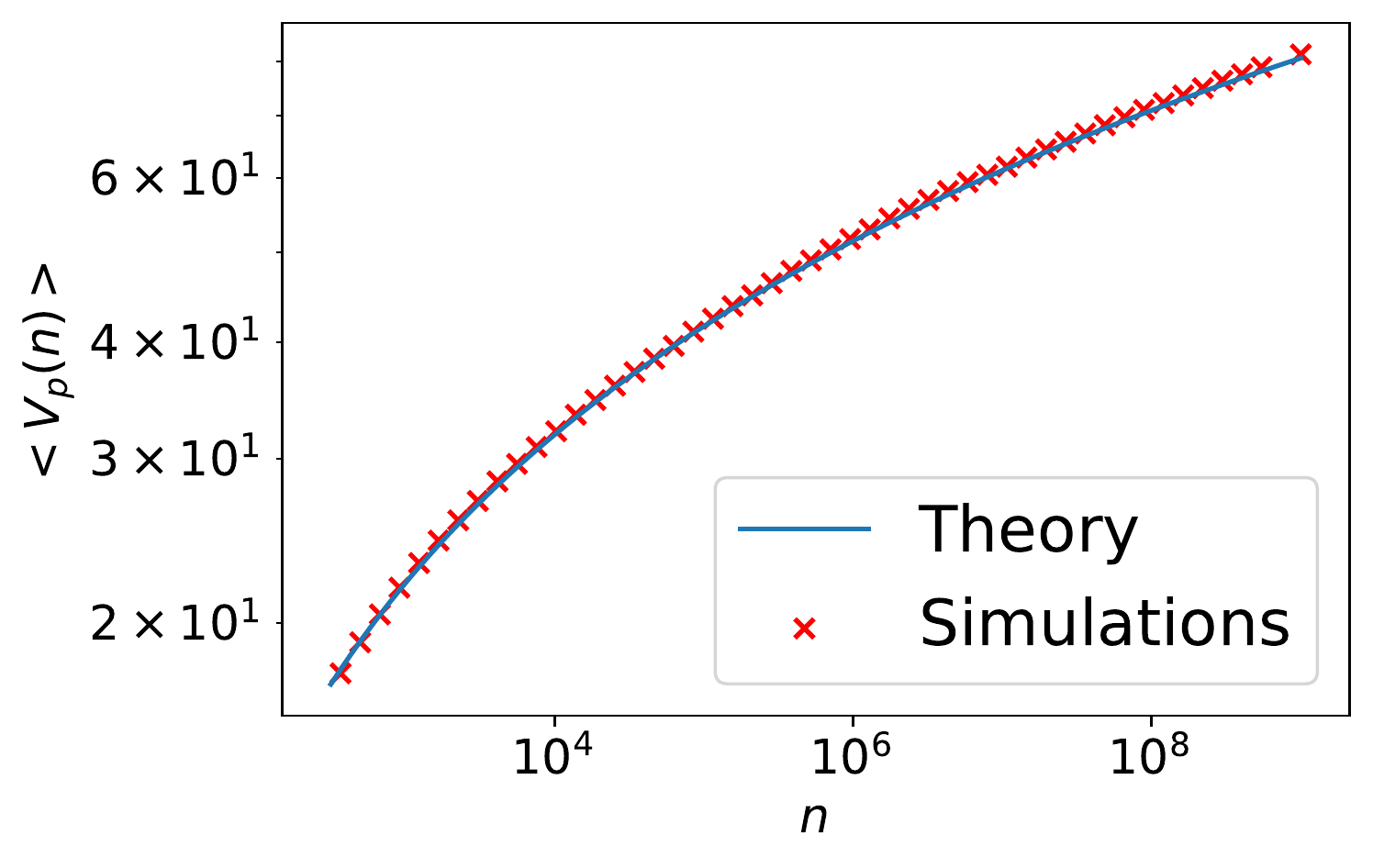}
\includegraphics[width=0.5\textwidth]{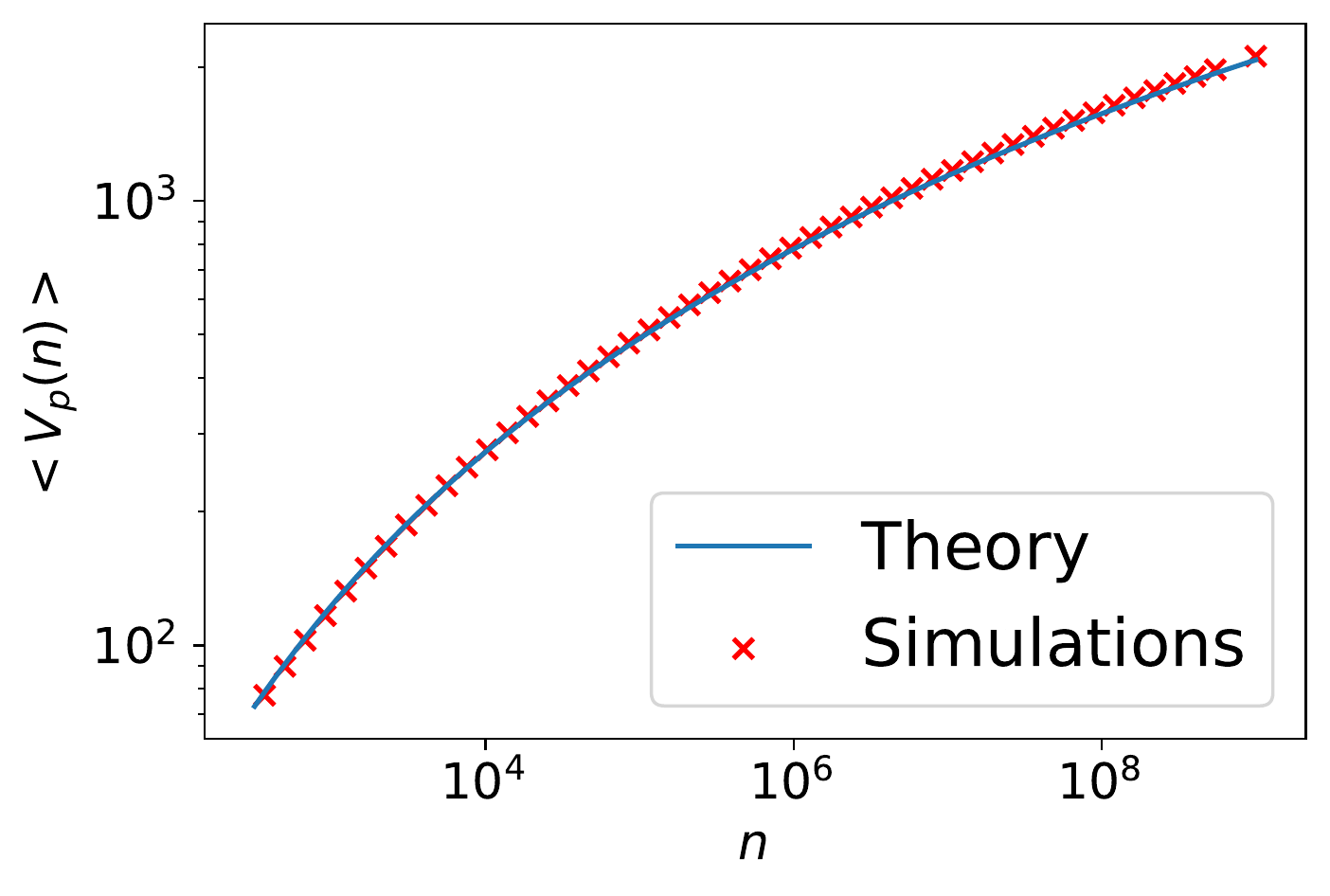}
\includegraphics[width=0.5\textwidth]{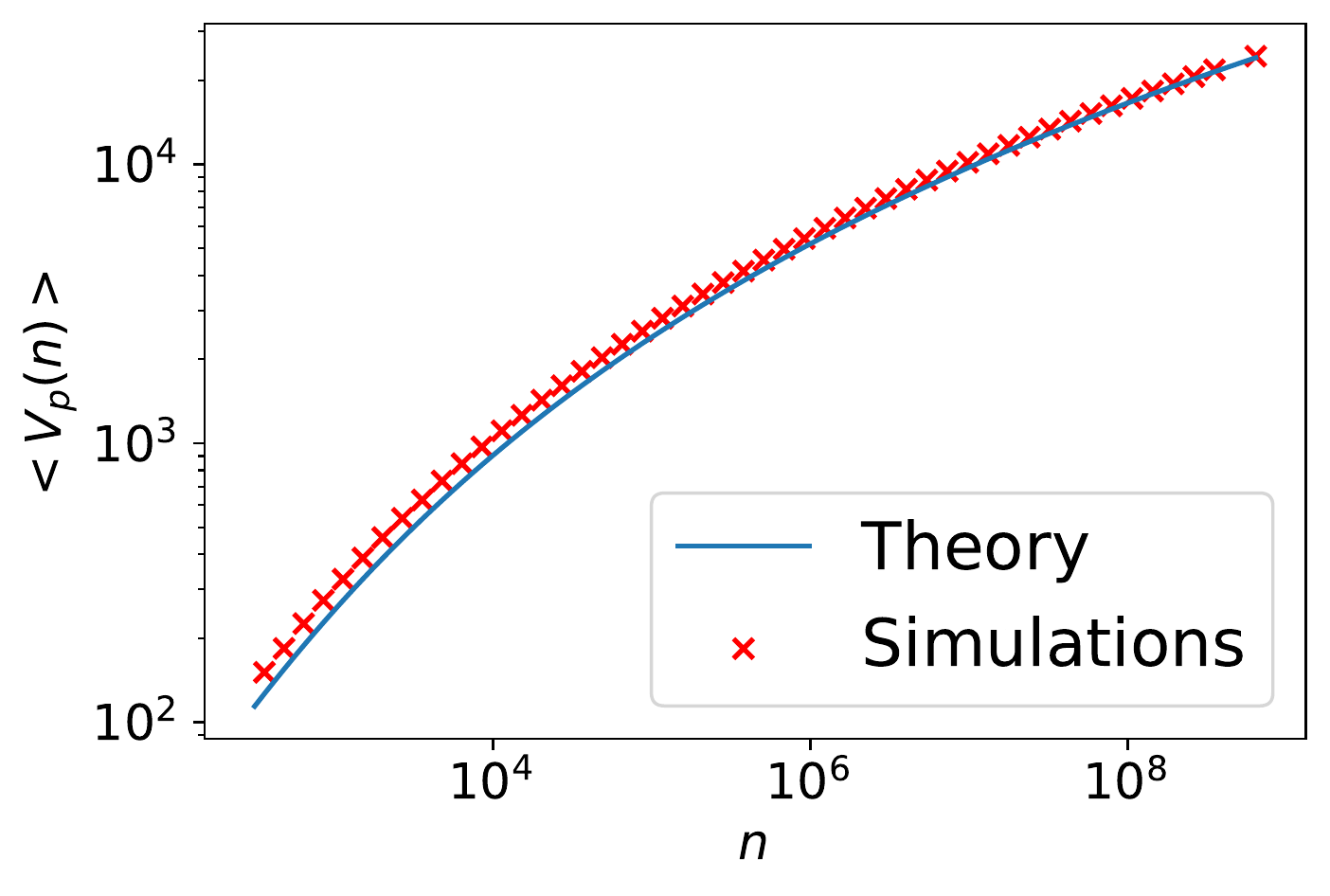}
\caption{The average number $\langle V_p(n)\rangle$ of distinct visited sites as a function of the number $n$ of steps for $d=1$ (top), $d=2$ (center), and $d=3$ (bottom). The red crosses correspond to numerical simulations performed averaging over $10^3$ trajectories. The continuous blue lines correspond to the asymptotic result given in Eq.~\ref{Vp_full}. 
\label{fig:V}}
\end{figure}

In the rest of this section, we analyze the expression in Eq.~\ref{identity_Vp} and extract the late-time behavior of $\langle V_p(n)\rangle$. As shown in Section \ref{sec:p0}, the large-$n$ limit corresponds to the limit $z\to 1$. Introducing the variable $s=1-z$, we obtain
\begin{equation}
\int_{0}^{\infty}dn~\langle V_p(n)\rangle e^{-sn}\approx\frac{s+p}{s}\sum_{\vb{X}}\left[p+s\frac{\tilde{G}_0(\vb{0},\vb{0},(1-s)(1-p))}{\tilde{G}_0(\vb{0},\vb{X},(1-s)(1-p))}\right]^{-1}\,.
\label{identity_Vp2}
\end{equation}
Moreover, when $p$ is fixed and $s\to 0$, we get
\begin{equation}
\int_{0}^{\infty}dn~\langle V_p(n)\rangle e^{-sn}\approx\frac{p}{s}\sum_{\vb{X}}\left[p+s\frac{\tilde{G}_0(\vb{0},\vb{0},1-p)}{\tilde{G}_0(\vb{0},\vb{X},1-p)}\right]^{-1}\,.
\label{identity_Vp3}
\end{equation}
For small $s$, the sum on the right-hand side is dominated by large values of $|\vb{X}|$. It is therefore instructive to investigate the asymptotic behavior of $\tilde{G}_0(\vb{0},\vb{X},1-p)$ for large $|\vb{X}|$, which is given by \ref{app:asympt_G0}
\begin{equation}
\tilde{G}_0(\vb{0},\vb{X},1-p)\approx \frac{1}{2p(4\pi)^{d/2}}\left(\frac{1-p}{8dp}\right)^{-(d+2)/4}|\vb{X}|^{1-d/2}K_{1-d/2}\left(\sqrt{\frac{2dp}{1-p}}|\vb{X}|\right)\,,\label{G0_asympX_bessel}
\end{equation}
where $K_{\nu}(z)$ is the modified Bessel function of the second kind and $|\vb{X}|$ indicates the Euclidean norm of $\vb{X}$. Moreover, using the asymptotic expression for $z\gg 1$ \cite{Bessel}
\begin{equation}
K_{\nu}(z)\approx\sqrt{\frac{\pi}{2z}}e^{-z}\,,
\label{K_asymp}
\end{equation}
we obtain
\begin{equation}
\tilde{G}_0(\vb{0},\vb{X},1-p)\approx g_d(p) |\vb{X}|^{(1-d)/2} e^{-\sqrt{2dp/(1-p)}|\vb{X}|}\,,\label{G0_asympX}
\end{equation}
where 
\begin{equation}
g_d(p)=\frac{\sqrt{\pi}}{2p(4\pi)^{d/2}}\left(\frac{8dp}{1-p}\right)^{(d+1)/4}\,.
\label{gp}
\end{equation}
Plugging this expansion into the expression in Eq.~\ref{identity_Vp3} and replacing the sum by an integral, we obtain
\begin{equation}
\int_{0}^{\infty}dn~\langle V_p(n)\rangle e^{-sn}\approx\frac1s\int d^d\vb{X}~\left[1+s C_d(p) |X|^{(d-1)/2}\exp\left(\sqrt{\frac{2dp}{1-p}}|X|\right)\right]^{-1}\,,
\label{identity_Vp4}
\end{equation}
where 
\begin{equation}
C_d(p)=\frac{\tilde{G}_0(\vb{0},\vb{0},1-p)}{pg_d(p)}=\frac{2(4\pi)^{d/2}}{\sqrt{\pi}}\left(\frac{1-p}{8dp}\right)^{(d+1)/4}\tilde{G}_0(\vb{0},\vb{0},1-p)\,,
\label{C_dp}
\end{equation}
and $\tilde{G}_0(\vb{0},\vb{0},1-p)$ is given in Eq.~\ref{eq:Gz0_expression}. Passing to spherical coordinates of integration, we obtain
\begin{equation}
\int_{0}^{\infty}dn~\langle V_p(n)\rangle e^{-sn}\approx \frac{\mathcal{A}(d)}{s}\int_{0}^{\infty} dR~R^{d-1}\left[1+s C_d(p) R^{(d-1)/2}\exp\left(\sqrt{\frac{2dp}{1-p}}R\right)\right]^{-1}\,,
\label{identity_Vp5}
\end{equation}
where $\mathcal{A}(d)$ is the surface of the $d$-dimensional hypersphere, given in Eq.~\ref{eq:def_Ad}. We identify the poles $s=0$ and 
\begin{equation}
s=-\frac{1}{C_d(p)R^{(d-1)/2}\exp\left(\sqrt{\frac{2dp}{1-p}}R\right)}
\end{equation}
of the expression on the right-hand side of Eq.~\ref{identity_Vp5} and we evaluate the corresponding residues to obtain 
\begin{equation}
\langle V_p(n)\rangle \approx \mathcal{A}(d)\int_{0}^{\infty} dR~R^{d-1}\left[1-\exp\left(-\frac{n}{C_d(p) R^{(d-1)/2}\exp\left(\sqrt{\frac{2dp}{1-p}}R\right)}\right)\right]\,,
\label{Vp_full}
\end{equation}
which is valid for large $n$. The constant $C_d(p)$ is given in Eq.~\ref{C_dp} and can be evaluated numerically at fixed $d$ and $p$. This asymptotic result is shown in Fig.~\ref{fig:V} for $d=1,2,3$ and is in excellent agreement with numerical simulations performed for $p=0.1$.

We are now interested in extracting the leading-order behavior of $\langle V_p(n)\rangle$ for large $n$. As we will show, this leading order will be completely independent of the constant $C_d(p)$. To proceed, we focus on the small-$s$ behavior of the integrand in Eq.~\ref{identity_Vp5}, yielding
\begin{equation}
\frac{1}{1+s C_d(p) R^{(d-1)/2}\exp\left(\sqrt{\frac{2dp}{1-p}}R\right)}=\frac{1}{1+\exp\left(\sqrt{\frac{2dp}{1-p}}R+\log(s)+\ldots\right)}\,.
\end{equation}
For small $s$, this can be approximated as
\begin{equation}
\frac{1}{1+s C_d(p) R^{(d-1)/2}\exp\left(\sqrt{\frac{2dp}{1-p}}R\right)}\approx\begin{cases}
1\,,&\text{ for }R<-\log(s)\sqrt{(1-p)/(2dp)}\,,\\
0\,,&\text{ for }R>-\log(s)\sqrt{(1-p)/(2dp)}\,.
\end{cases}
\end{equation}
Indeed, the left-hand side assumes the form of the Fermi function from quantum mechanics and this approximation corresponds to a low-temperature expansion. With this approximation, Eq.~\ref{identity_Vp5} becomes
\begin{equation}
\int_{0}^{\infty}dn~\langle V_p(n)\rangle e^{-sn}\approx \frac{\mathcal{A}(d)}{s}\int_{0}^{-\log(s)\sqrt{(1-p)/(2dp)}} dR~R^{d-1}=\frac{\mathcal{A}(d)}{d}\frac{1}{s}\left[\sqrt{\frac{1-p}{2dp}}\log(\frac1s)\right]^d\,,
\label{identity_Vp6}
\end{equation}
Finally, inverting the Laplace transform, we obtain that, for $p>0$ and $n\gg1$,
\begin{equation}
\langle V_p(n)\rangle\approx\frac{\mathcal{A}(d)}{d}\left(\frac{1-p}{2dp}\right)^{d/2}\left[\log(n)\right]^d+\mathcal{O}\left[\left(\log(n)\right)^{d-1}\right]\,,
\label{Vp_sec}
\end{equation}
which is the main result of this paper. Notably, at variance with the case $p=0$, we obtain the same asymptotic result for any $d>0$ meaning that no recurrence-transience transition is observed. Interestingly, this leading-order result is completely independent of the constant $C_d(p)$, which depends on the precise details of the lattice. Therefore, we expect our result in Eq.~\ref{Vp_sec} to be universal, i.e., valid for any regular lattice in $d$ dimensions. 

It is interesting to compare our results on the number of distinct sites in Eq.~\ref{Vp_sec} with those of Ref.~\cite{MajumdarConvex} on the convex hull of a two-dimensional resetting BM. Note that the set of visited sites is always a subset of the convex hull of a process. Indeed, the convex hull can in principle contain sites that are yet to be visited but every visited site belongs by definition to the convex hull. For $d=1$ the visited sites and the convex hull coincide since there can be no holes in the set of visited sites. This is not true in general for $d>1$, where non-visited sites are typically present in the convex hull of the process. For instance, for a two-dimensional RW without resetting, we have shown that the average number of visited sites grows as $\langle V_0(n)\rangle\approx \pi n/\log(n)$ in the limit of many steps. For the same process, it is well known that the number $A(n)$ of sites in the convex hull grows on average as $\langle A(n)\rangle\approx \pi n/4$ for large $n$ \cite{MajumdarConvex}. As a consequence, the fraction of sites contained in the convex hull that are actually visited by the process decreases as $\langle V_0(n)\rangle/\langle A(n)\rangle\approx 4/\log(n)$ for large $n$, meaning that very few sites in the convex hull are actually visited.

To compare our result for $d=2$ with those of Ref.~\cite{MajumdarConvex} we need to consider the continuous time limit $p\to 0$, $n\to \infty$ with $pn$ fixed. In this limit, our result in Eq.~\ref{Vp_sec} becomes
\begin{equation}
\langle V_p(n)\rangle\approx\frac{\pi}{4p}\left[\log(n)\right]^2+\mathcal{O}(\log(n))\,.
\end{equation}
In Ref.~\cite{MajumdarConvex}, the area $A(n)$ of the convex hull was shown to grow as
\begin{equation}
\langle A(n)\rangle\approx\frac{\pi}{4p}\left[\log(n)\right]^2+\mathcal{O}(\log(n))\,,
\end{equation}
for large $n$. In other words, $\langle V_p(n)\rangle$ and $\langle A(n)\rangle$ have the same asymptotic behavior at leading order. Therefore, for large $n$ almost every site in the convex hull of a resetting random walker has been visited at least once.

\subsection{Intermediate scaling regime}
\label{sec:scaling}

In the sections above we have shown that the average number of distinct visited sites has very different behaviors in the cases of random walks with resetting, where $\langle V_p(n)\rangle$ grows logarithmically in $n$, and without resetting, where the growth is polynomial in $n$. It is therefore relevant to ask how the crossover between these two regimes occurs when $p$ is small. It is useful to consider the cases of $d<2$ and $d>2$ separately.

For $d<2$, it turns out that the crossover occurs in the scaling regime where $p\to 0$ and $n\to \infty$ with $pn$ fixed. In terms of the Laplace variable $s$, this corresponds to the limit $s,p\to 0$ with $s/p$ fixed. Going back to Eq.~\ref{identity_Vp2} and considering this limit, we obtain
\begin{equation}
\int_{0}^{\infty}dn~\langle V_p(n)\rangle e^{-sn}\approx\frac{s+p}{s}\sum_{\vb{X}}\left[p+s\frac{\tilde{G}_0(\vb{0},\vb{0},1-s-p)}{\tilde{G}_0(\vb{0},\vb{X},1-s-p)}\right]^{-1}\,.
\label{identity_Vp_scaling}
\end{equation}
Using the asymptotic expressions for $\tilde{G}_0(\vb{0},\vb{0},1\!-\!s\!-\!p)$ and $\tilde{G}_0(\vb{0},\vb{X},1\!-\!s\!-\!p)$, respectively given in Eqs.~\ref{G0_asymp} and \ref{G0_asympX_bessel}, we obtain
\begin{equation}
\int_{0}^{\infty}dn~\langle V_p(n)\rangle e^{-sn}\approx\frac{s+p}{s}\sum_{\vb{X}}\left[p+\frac{ A_d \left(\sqrt{p+s}|\vb{X}|\right)^{(d-2)/2}}{2^{(2-d)/4}\pi^{-d/2}d^{(d+2)/4} K_{1-\frac{d}{2}}\left(\sqrt{2d(p+s)}|\vb{X}|\right)}\,s\right]^{-1}\,,
\label{identity_Vp_scaling2}
\end{equation}
where $A_d$ is given in Eq.~\ref{A_d}. Substituting the sum by an integral over spherical coordinates yields
\begin{align}
&\int_{0}^{\infty}dn~\langle V_p(n)\rangle e^{-sn}\\
&\approx  \frac{s+p}{s}\mathcal{A}(d)\int_{0}^{\infty}dR~R^{d-1}\left[p+\frac{ A_d\left(\sqrt{p+s}R\right)^{(d-2)/2}}{2^{(2-d)/4}\pi^{-d/2}d^{(d+2)/4}K_{1-\frac{d}{2}}\left(\sqrt{2d(p+s)}R\right)}\,s\right]^{-1}\,.\nonumber
\end{align}
Performing the change of variables $R\to y=\sqrt{2d(p+s)}R$, we get
\begin{equation}
\int_{0}^{\infty}dn~\langle V_p(n)\rangle e^{-sn}\approx\frac{(s+p)^{1-d/2}}{s}\frac{\mathcal{A}(d)}{(2d)^{d/2}}\int_{0}^{\infty}dy~y^{d-1}\left[p+\frac{\pi^{d/2}A_d  y^{(d-2)/2}}{d^{d/2} K_{1-\frac{d}{2}}(y)}\,s\right]^{-1}\,.
\end{equation}
Inverting the Laplace transform formally, we obtain
\begin{equation}\langle V_p(n)\rangle \approx\frac{\mathcal{A}(d)}{(2d)^{d/2}} \int_{\Gamma}\frac{ds}{2\pi i}e^{sn}\frac{(s+p)^{1-d/2}}{s}\int_{0}^{\infty}dy~y^{d-1}\left[p+\frac{\pi^{d/2}A_d  y^{(d-2)/2}}{d^{d/2} K_{1-\frac{d}{2}}(y)}\,s\right]^{-1}\,,
\end{equation}
where the integral over $s$ runs over the Bromwich contour in the complex-$s$ plane. Finally, this expression can be written in the scaling form, valid for $p\to 0$ and $n\to \infty$ with $pn$ fixed,
\begin{equation}
\langle V_p(n)\rangle \approx\frac{\mathcal{A}(d)}{(2dp)^{d/2}}F_d(pn)\,,
\label{scal_d<2}
\end{equation}
where
\begin{equation}
F_d(z)= \int_{\Gamma}\frac{dq}{2\pi i}e^{qz}\frac{(1+q)^{1-d/2}}{q}\int_{0}^{\infty}dy~y^{d-1}\left[1+\frac{\pi^{d/2}A_d  y^{(d-2)/2}}{d^{d/2} K_{1-\frac{d}{2}}(y)}~q\right]^{-1}\,.\label{fd_d<2}
\end{equation}

\begin{figure}
\centering
\includegraphics[width=0.5\textwidth]{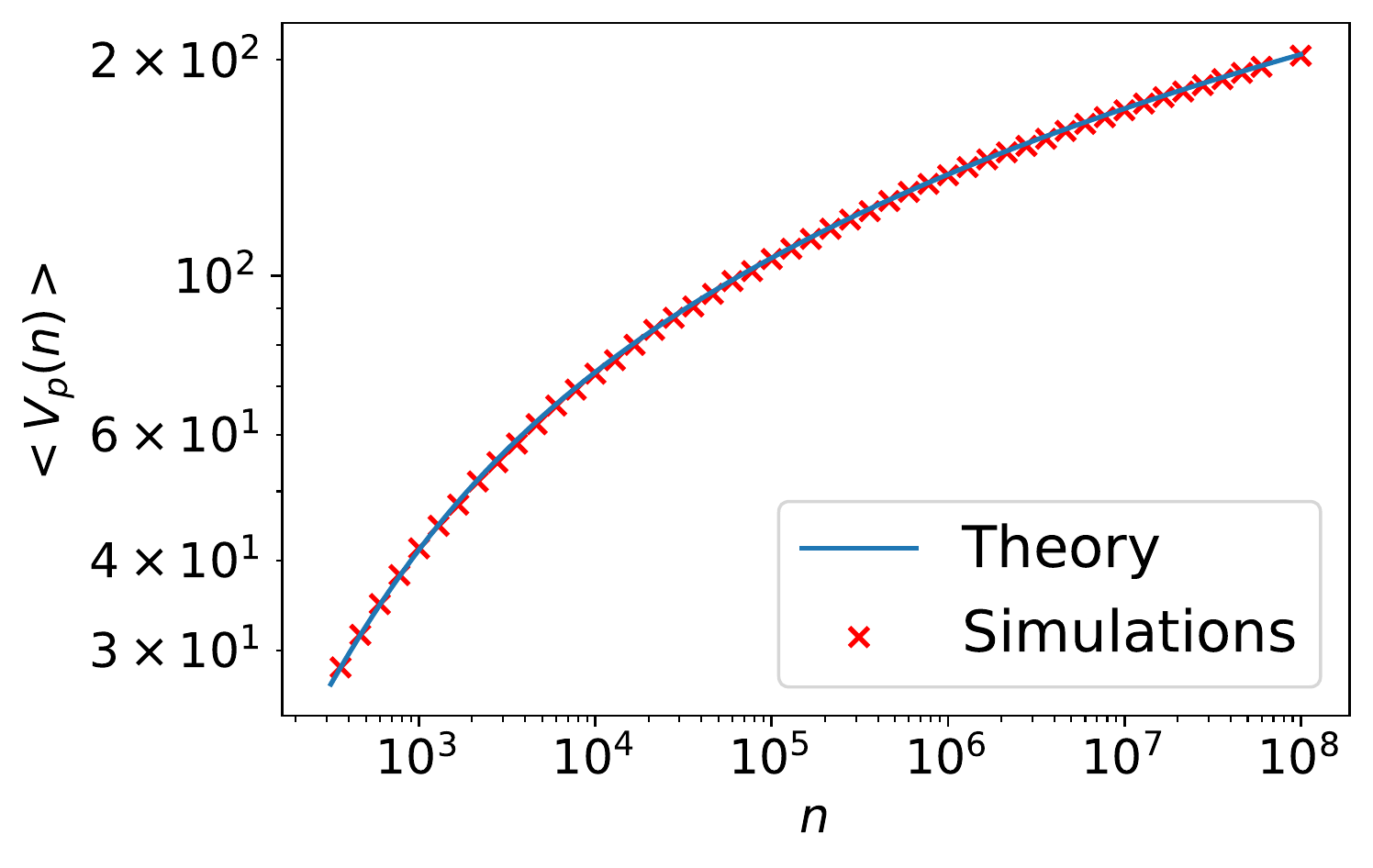}
\includegraphics[width=0.5\textwidth]{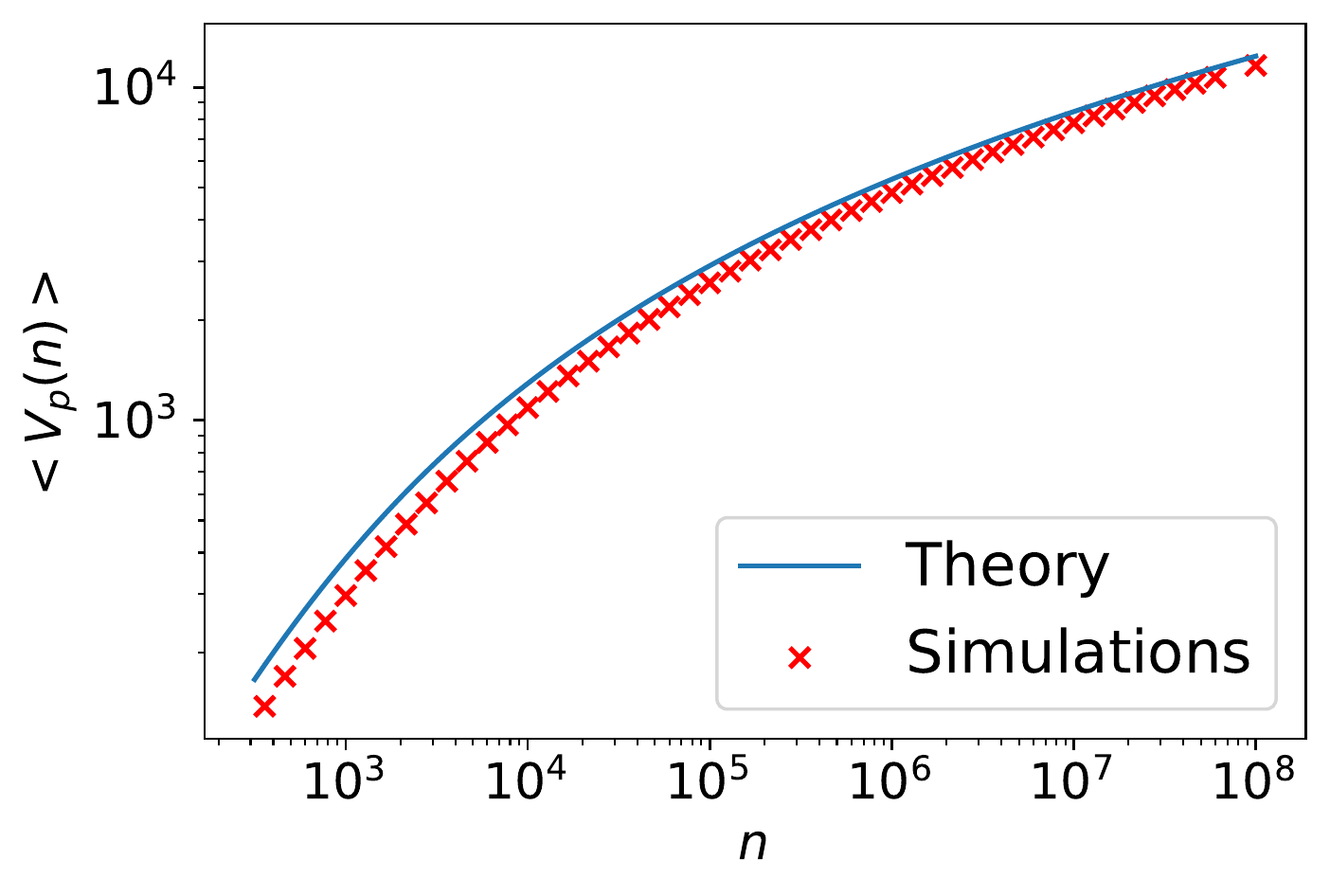}
\includegraphics[width=0.5\textwidth]{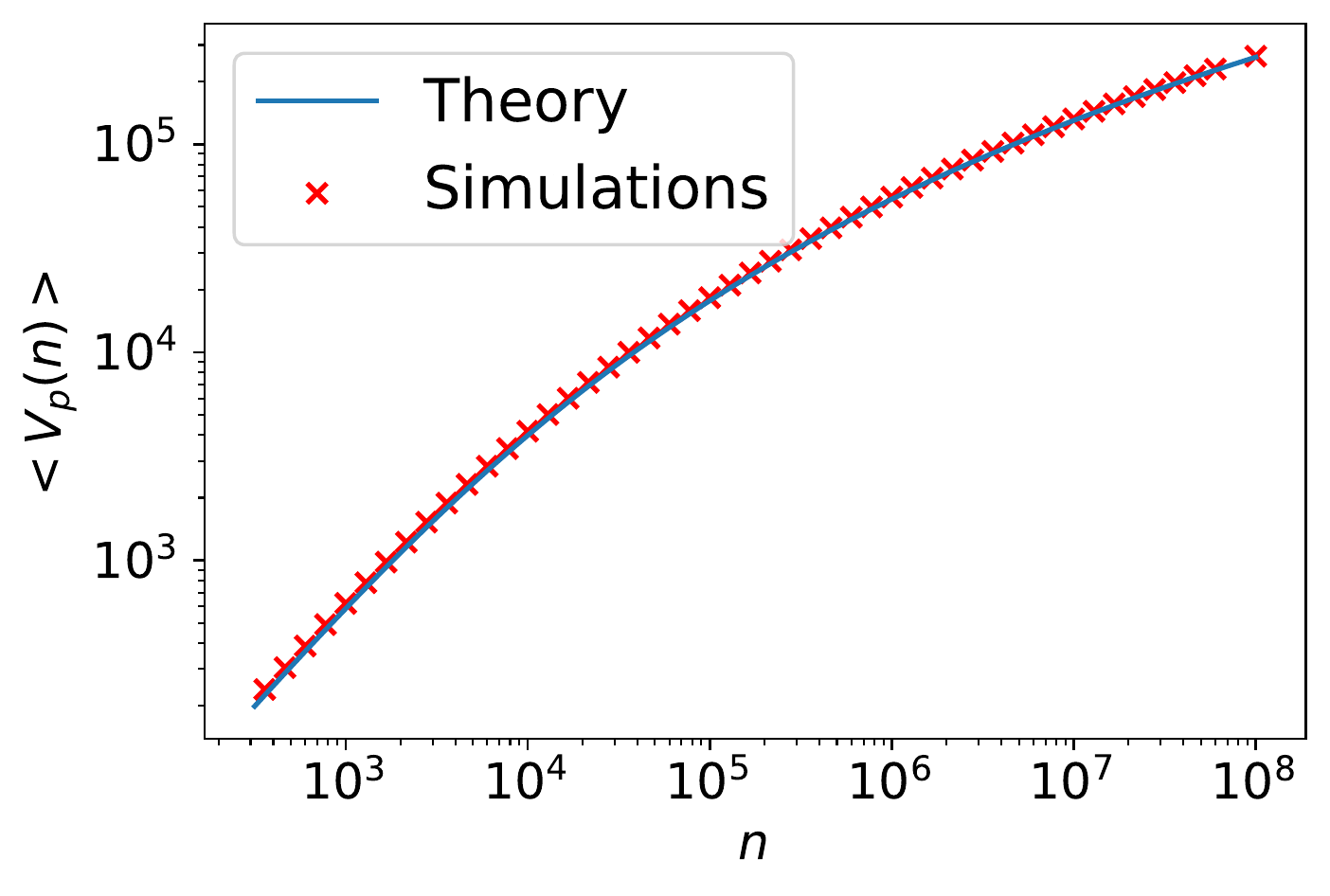}
\caption{The average number $\langle V_p(n)\rangle$ of distinct visited sites as a function of the number $n$ of steps for $d=1$ (top), $d=2$ (center), and $d=3$ (bottom) in the scaling regime $p\to 0$, $n\to\infty$. The continuous blue line corresponds to the analytical asymptotic result given in Eqs.~\ref{scal_d<2} and \ref{eq:f1z} for $d=1$, in Eqs.~\ref{Vp_scaling_d2} and \ref{f2} for $d=2$, and in Eqs.~\ref{scal_d>2} and \ref{eq:f3z} for $d=3$. The red symbols are the results of numerical simulations performed with $p=0.01$. Note that in $d=2$ the agreement between theory and simulation is less precise due to logarithmic corrections.
\label{fig:V_scaling}}
\end{figure}

The asymptotic behaviors of this scaling function can be easily obtained and one gets
\begin{equation}
F_d(z)\approx\begin{cases}
\frac{(2d)^{d/2}}{\mathcal{A}(d) A_d \Gamma(1+d/2)}z^{d/2}\,,&\text{ for }z\to 0\,,\\
\\
\frac{1}{ d}(\log z)^{d}\,,&\text{ for }z\to \infty\,.
\end{cases}
\end{equation}
Correspondingly, the asymptotic behaviors of $\langle V_p(n)\rangle$ read
\begin{equation}
\langle V_p(n)\rangle \approx\begin{cases}
\frac{1}{A_d \Gamma(1+d/2)}n^{d/2}\,,&\text{ for }n\ll 1/p\,,\\
\\
\frac{\mathcal{A}(d)}{ (2p)^{d/2} d^{1+d/2}}[\log (pn)]^{d}\,,&\text{ for }n\gg 1/p\,,
\label{VP_asymp_d<2}
\end{cases}
\end{equation}
where $\mathcal{A}(d)$ and $A_d$ are respectively given in Eqs.~\ref{eq:def_Ad} and \ref{A_d}.
Thus, the scaling form in Eq.~\ref{scal_d<2} correctly interpolates between the cases cases $p=0$ (see Eq.~\ref{V0_d<2}) and $p>0$ (see Eq.~\ref{Vp_sec}). This can be simply understood as follows. Since $1/p$ is the typical time at which the first resetting occurs, when $n\ll 1/p$ the random walker has hardly undergone any resetting. Therefore, $\langle V_p(n)\rangle$ grows as in the case without resetting. On the other hand, for $n\gg 1/p$, many resetting events have already occurred and one obtains the result derived for resetting random walks. For $d=1$, the expression in Eq.~\ref{fd_d<2} becomes
\begin{equation}
F_1(z)= \int_{\Gamma}\frac{dq}{2\pi i}e^{qz}\frac{\sqrt{1+q}}{q}\int_{0}^{\infty}dy~\frac{1}{1+q~e^{y}}\,,
\end{equation}
where we have used $A_1=1/\sqrt{2}$ (see Eq.~\ref{A_d}) and $K_{1/2}(y)=e^{-y}\sqrt{\pi/(2y)}$ \cite{Bessel}. Computing the integral over $y$, we find
\begin{equation}
F_1(z)= \int_{\Gamma}\frac{dq}{2\pi i}e^{qz}\frac{\sqrt{1+q}}{q}\log(\frac{q+1}{q})\,.
\label{lapl_F1}
\end{equation}
This Laplace transform can be inverted \ref{app:F_1_inversion} and one obtains
\begin{equation}
F_1(z)=\int_{0}^{z}\frac{dx}{x}\left(1-e^{-x}\right)\left[\frac{e^{-(z-x)}}{\sqrt{\pi(z-x)}}+\erf(\sqrt{z-x})\right]\,.
\label{eq:f1z}
\end{equation}
Interestingly, this scaling function $F_1(z)$ also describes the average maximum of resetting BM in one dimension \cite{MajumdarConvex}. This connection with resetting BM is explored in detail in Section \ref{sec:one-dim}.

We next focus on the case $d>2$. It turns out that in this case the correct scaling regime is obtained for $n\to\infty$ and $p\to 0$ with $z=np^{d/2}$ fixed. In terms of Laplace variables, this corresponds to the limit $s,p\to 0$ with $s/p^{d/2}$ fixed. Note that $s\ll p$ since $d>2$. Plugging the asymptotic behavior of $\tilde{G}_0(\vb{0},\vb{0},1-s-p)$ and $\tilde{G}_0(\vb{0},\vb{X},1-s-p)$, given in Eqs.~\ref{G0_asymp} and \ref{G0_asympX_bessel}, into Eq.~\ref{identity_Vp_scaling}, we obtain
\begin{equation}
\int_{0}^{\infty}dn~\langle V_p(n)\rangle e^{-sn}\approx\frac{1}{s}\sum_{\vb{X}}\left[1+\frac{s}{p^{d/2}}\frac{ \pi^{d/2}B_d\left(\sqrt{2dp}|\vb{X}|\right)^{(d-2)/2}}{d^{d/2} K_{1-\frac{d}{2}}\left(\sqrt{2dp}|\vb{X}|\right)}\right]^{-1}\,,
\end{equation}
where $B_d=\tilde{G}(\vb{0},\vb{0},1)$. Approximating the sum by an integral over spherical coordinates, we get
\begin{equation}
\int_{0}^{\infty}dn~\langle V_p(n)\rangle e^{-sn}\approx\frac{\mathcal{A}(d)}{s}\int_{0}^{\infty}dR~R^{d-1}\left[1+\frac{s}{p^{d/2}}\frac{ \pi^{d/2}B_d\left(\sqrt{2dp}R\right)^{(d-2)/2}}{d^{d/2} K_{1-\frac{d}{2}}\left(\sqrt{2dp}R\right)}\right]^{-1}\,.
\end{equation}
Performing the change of variable $R\to y=\sqrt{2dp}R$ yields
\begin{equation}
\int_{0}^{\infty}dn~\langle V_p(n)\rangle e^{-sn}\approx\frac{\mathcal{A}(d)}{(2dp)^{d/2}s}\int_{0}^{\infty}dy~y^{d-1}\left[1+\frac{s}{p^{d/2}}\frac{ \pi^{d/2}B_d y^{(d-2)/2}}{d^{d/2} K_{1-\frac{d}{2}}\left(y\right)}\right]^{-1}\,.
\end{equation}
The expression on the right-hand side has poles at $s\!=\!0$ and 
\begin{equation}s\!=\!-\frac{(pd)^{d/2}K_{1-d/2}(y)}{\pi^{d/2}B_d y^{(d-2)/2}}\,.
\end{equation}
Evaluating the residues at these poles, we get
\begin{equation}
\langle V_p(n)\rangle\approx \frac{\mathcal{A}(d)}{(2dp)^{d/2}} F_d(p^{d/2} n)
\label{scal_d>2}
\end{equation}
where, for $d>2$,
\begin{equation}
F_d(z)=\int_{0}^{\infty}dy~y^{d-1}\left[1-\exp\left(-\frac{d^{d/2} K_{1-\frac{d}{2}}\left(y\right)}{ \pi^{d/2}B_d y^{(d-2)/2}}z\right)\right]\,.
\label{fd_d>2}
\end{equation}
For instance, for $d=3$ we obtain
\begin{equation}
F_3(z)=\int_{0}^{\infty}dy~y^{2}\left[1-\exp\left(-\frac{3^{3/2} K_{-\frac{1}{2}}\left(y\right)}{ \pi^{3/2}B_3 \sqrt{y}}z\right)\right]\,.
\label{eq:f3z}
\end{equation}
where $B_3=1.516\ldots$ and $K_{-1/2}(y)=K_{1/2}(y)=e^{-y}\sqrt{\pi/(2y)}$.

For $d>2$, the asymptotic behaviors of $F_d(z)$ for small and large $z$ are given by
\begin{equation}
F_d(z)\approx\begin{cases}
\frac{(2d)^{d/2}}{\mathcal{A}(d) B_d}z\,,&\text{ for }z\to 0\,,\\
\\
\frac{1}{ d}(\log z)^{d}\,,&\text{ for }z\to \infty\,.
\end{cases}
\end{equation}
From this expansions, we obtain
\begin{equation}
\langle V_p(n)\rangle\approx\begin{cases}
\frac{1}{B_d}n\,,&\text{ for }n\ll 1/p^{d/2}\,,\\
\\
\frac{\mathcal{A}(d)}{ (2p)^{d/2} d^{1+d/2}}[\log (pn)]^{d}\,,&\text{ for }n\gg 1/p^{d/2}\,.
\label{VP_asymp_d>2}
\end{cases}
\end{equation}
Therefore, the scaling form above correctly reproduces the two limit cases $p=0$ and $p>0$, given in Eqs.~\ref{V0_d>2} and \ref{Vp_sec}.

In the case $d=2$, following the derivation as in the previous cases, we obtain that the scaling regime to be considered for $n\to\infty$ and $p\to 0$ with $z=pn/\log(1/p)$ fixed. In this limit, we show
\begin{equation}
\langle V_p(n)\rangle\approx\frac{\pi}{2p}F_2\left(\frac{pn}{\log(1/p)}\right)\,,
\label{Vp_scaling_d2}
\end{equation}
where 
\begin{equation}
F_2(z)=\int_{0}^{\infty}dy~y\left[1-\exp(-2 K_0(y)z)\right]\,.
\label{f2}
\end{equation}
This scaling function has the asymptotic behaviors
\begin{equation}
F_2(z)\approx\begin{cases}
2z\,,&\text{ for }z\to 0\,,\\
\\
\frac{1}{2}(\log z)^{2}\,,&\text{ for }z\to \infty\,,
\end{cases}
\end{equation}
correctly reproducing the results in Eqs.~\ref{V0_d=2} and \ref{Vp_sec}. The results for $\langle V_p(n)\rangle$ in the scaling regime are shown in Fig.~\ref{fig:V_scaling} for $d=1,2,3$ and are in perfect agreement with numerical simulations.

Finally, we provide a simple argument as to why the crossover between the two regimes occurs after $n\sim 1/p$ steps for $d<2$ and $n\sim 1/p^{d/2}$ steps for $d>2$. We recall that the steady state distribution of a $d$-dimensional resetting RW is given by \cite{EM14}
\begin{equation}
P_{\rm st}(\vb{X})=\left(\frac{p}{\pi}\right)^{d/2}\left(\sqrt{2p}|\vb{X}|\right)^{1-d/2}K_{1-d/2}\left(\sqrt{2p}|\vb{X}|\right)\,.
\end{equation}
Thus, in the stationary state, the walker will be typically located within a distance of order $p^{-1/2}$ from the resetting site $\vb{X}=0$. In other words, the sites typically visited by the walker in the steady state are located in a typical hypersphere of radius $R_{\rm typ}\sim p^{-1/2}$. From time to time, the walker will then venture outside of this set of typical sites to discover new sites. However, visiting new sites becomes harder with time, leading to the slow logarithmic growth of $\langle V_p(n)\rangle$.

Therefore, the logarithmic growth will appear once two conditions have been met: (i) several resettings have occurred, inducing a steady state, and (ii) the walker has visited every site within a distance $R_{\rm typ}$ from the origin. These conditions induce two different timescales in the process. Indeed, condition (i) will surely be satisfied for $n\gg n_1^*$, where $n_1^*\sim p^{-1}$ is the typical time between successive reset events. On the other hand, for $d>2$ the second condition requires at least $n_2^*\sim R_{\rm typ}^d$ steps, since $V_0(n)\sim n$. For $d<2$, the number of visited sites without resetting grows slower as $V_0(n)\sim n^{d/2}$ and thus the second condition will be satisfied only for $n\gg n_2^*\sim R_{\rm typ}^2$. Therefore, condition (ii) is satisfied for $n\gg n_2^*$, where $n_2^*\sim p^{-d/2}$ for $d>2$ and $n_2^*\sim 1/p$ for $d<2$. Since both conditions are required, the crossover will occur at $n^*\sim \max\left[n_1^*,n_2^*\right]$. For $d<2$, both timescales $n_1^*$ and $n_2^*$ scale as $1/p$ and therefore $n^*\sim 1/p$. Conversely, for $d>2$, the typical hypersphere will be fully visited after several resetting events (i.e., $n_1^*\ll n_2^*$), corresponding to $n^*\sim p^{-d/2}$.

\section{Resetting random walk in one dimension}

\label{sec:one-dim}

\begin{figure}
\begin{tikzpicture}[xscale=1]

\draw [-> , help lines,thick] (-1,0) -- (11, 0);
\draw [help lines,->,thick] (0, -4) -- (0, 4);

\node [-,black] at (0,0) {\textbullet};
\draw [blue, thick] (0,0) -- (1,1);
\node [-,black] at (1,1) {\textbullet};
\draw [blue, thick]  (1,1)-- (2,2);
\node [-,black] at (2,2) {\textbullet};
\draw [blue, thick]  (2,2)-- (3,1);
\node [-,black] at (3,1) {\textbullet};
\draw [blue, thick]   (3,1)--(4,2);
\node [-,black] at (4,2) {\textbullet};
\draw [->,red, thick]   (4,2)--(4.5,1);
\draw [-,red, thick]   (4.5,1)--(5,0);
\draw [red, thick]   (4,2)--(5,0);
\node [-,black] at (5,0) {\textbullet};
\draw [blue, thick]   (5,0)--(6,-1);
\node [-,black] at (6,-1) {\textbullet};
\draw [blue, thick]   (6,-1)--(7,-2);
\node [-,black] at (7,-2) {\textbullet};
\draw [blue, thick]   (7,-2)--(8,-3);
\node [-,black] at (8,-3) {\textbullet};
\draw [->,red, thick]   (8,-3)--(8.5,-1.5);
\draw [red, thick]   (8.5,-1.5)--(9,0);
\node [-,black] at (9,0) {\textbullet};
\draw [blue, thick]   (9,0)--(10,-1);
\node [-,black] at (10,-1) {\textbullet};

\draw [black, dashed]   (2,2)--(0,2);
\draw [black, dashed]   (8,-3)--(0,-3);
\draw [black, dashed]   (10,-1)--(10,0);

\node[thick] at (-0.5, 2) {$M(t)$};
\node[thick] at (-0.5, -3) {$m(t)$};

\node[thick] at (0.5, -0.5) {$\Delta t$};

\draw [<->,black, thick]   (1.2,0)--(1.2,1);
\draw [<->,black, thick]   (0,-0.2)--(1.1,-0.2);

\draw [<->,black, thick]   (-1.5,-3)--(-1.5,2);

\node[thick] at (0.5, -0.5) {$\Delta t$};

\node[thick] at (1.5, 0.5) {$a$};

\node[black,thick] at (-2, 0) {$L(t)$};

\node[thick] at (-0.5, 4) {$x(\tau)$};

\node[thick] at (11,-0.5) {$\tau$};
\node[thick] at (10,0.2) {$t$};
\end{tikzpicture}
\caption{Typical realization of a resetting random walk (RW) $x(\tau)$ up to time $t$ in one dimension. The walker moves in a lattice with spacings $a$ and at each discrete step the total time is incremented by $\Delta t$. At each step, the walker is reset to the origin with probability $p=r\Delta t$ (red arrows). The maximum and minimum of the RW up to time $t$ are denoted by $M(t)$ and $m(t)$, respectively. The span $L(t)$, defined as $L(t)=M(t)-m(t)$, is related to the number of visited sites by $V_p(n)=L(t)/a$, where $n=t/\Delta t$ is the total number of steps.  \label{fig:trajectory}}
\end{figure}
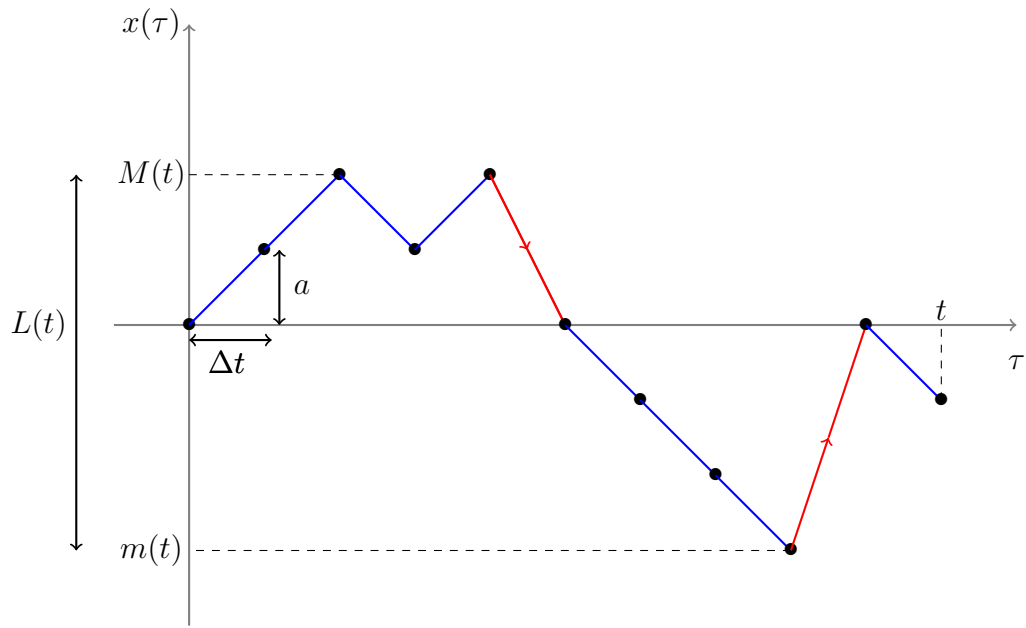

In one dimension the problem can be drastically simplified because the number of visited sites up to step $n$ can be written in terms of the maximum $M_n$ of the RW and the minimum $m_n$ as
\begin{equation}
V_n(p) = M_n - m_n \,,
\end{equation}
where $M_n=\max_{k\leq n} X_n$ and $m_n=\min_{k\leq n} X_n$.
Note that this relation between the extrema of the process and the number of visited sites is only valid in one dimension and cannot be easily generalized to higher dimensions. As we will see, this observation allows to go beyond the first moment of $V_n(p)$ and to compute its full distribution at late times. Since we are interested in the limit of many steps, we take directly the continuous-space and continuous-time limit. This can be done by introducing physical dimensions. We denote the lattice constant, i.e., the distance between two neighboring sites, by $a>0$. We also denote by $\Delta t>0$ the fixed time between two steps. Then the continuum limit corresponds to the limit $n\to\infty$, $a\to 0$, and $\Delta t\to 0$ with $D=a^2/(2\Delta t)$ fixed, where $D$ is the diffusion coefficient and $t=n\Delta t$ fixed, where $t$ is the total time. Similarly, as discussed in Section \ref{sec:scaling}, the resetting probability $p$ is rescaled as $p=r \Delta t$, where $r>0$ is the resetting rate. We will denote the resulting continuous-time stochastic process by $x(t)$, which is a one-dimensional resetting BM as a consequence of the central limit theorem. Finally, the continuous-time counterpart of the number of visited sites is the span $L(t)$, defined as
\begin{equation}
L(t)=M(t)-m(t)\,,
\label{eq:def_L}
\end{equation}
where $M(t)=\max_{\tau<t}x(\tau)$ and $m(t)=\min_{\tau<t}x(\tau)$. Thus, the span $L(t)$ and the number of visited sites $V_p(n)$ are related by (see Fig. \ref{fig:trajectory})
\begin{equation}
L(t)=\frac{1}{a}V_p(n)\,.
\end{equation}
Note that this continuum limit coincides with the scaling regime described in Section \ref{sec:scaling} for $d=1$.

It turns out that in one dimension it is possible to compute the full distribution of the span $L(t)$. To do this, we will first compute the joint PDF $P_r(M,m,t)$ of the maximum $M(t)$ and the minimum $m(t)$ up to time $t$ of the BM with resetting rate $r$. Then, we will use the definition in \ref{eq:def_L} to obtain the PDF of $L$. It is useful to notice that the joint cumulative distribution of $M$ and $m$ can be written in terms of the survival probability $Q_r(x_0,t|M,m)$ of a random walker with resetting moving in an interval $[m,M]$ with absorbing boundary conditions at $x=m$ and $x=M$ and starting from position $m<x_0<M$ at the initial time. Indeed, $Q_r(x_0,t|M,m)$ is defined as
\begin{equation}
Q_r(x_0,t|M,m)=\text{Prob.}(M(t)<M,m(t)>m)\,,
\end{equation}
corresponding to the probability that the walker does not leave the interval $[m,M]$ up to time $t$. It is useful to first consider the case $r = 0$, since it will provide the fundamental building block for the resetting case. The exact distribution of the span of a one-dimensional BM was first computed by Feller in 1951 \cite{F51}. We provide in the next section the derivation of this classical result, which will be useful to investigate the case of resetting.

\subsection{The case $r = 0$}

In the non-resetting case, the survival probability $Q_0(x, t | M, m)$ satisfies the backward Fokker-Planck equation
\begin{equation}\label{eq:max_fokker_planck}
        \pdv{Q_0(x, t | M, m)}{t} = D \pdv[2]{Q_0(x, t | M, m)}{x}.
\end{equation}
with absorbing boundary conditions
\begin{equation}
\begin{cases}
Q_0(x=M, t | M, m) = 0\,,\\
Q_0(x=m, t | M, m) = 0\,,
\end{cases}
\end{equation}
and initial condition
\begin{equation}\label{eq:init_condition}
Q_0(x, 0 | M, m) = 1\,.
\end{equation}
To solve this equation, it is useful to take a Laplace transform on both sides of Eq.~\ref{eq:max_fokker_planck}, yielding
\begin{equation}\label{eq:max_lapl_fokker_planck}
 s \tilde{Q_0}(x, s | M, m) - 1 = D \pdv[2]{\tilde{Q_0}(x, s| M, m)}{x},
\end{equation}
where
\begin{equation}
\tilde{Q_0}(x, s | M, m)=\int_{0}^{\infty}dt~Q_0(x, t | M, m)e^{-st}\,,
\end{equation}
and we have used the initial condition given in \ref{eq:init_condition}. The boundary conditions remain unchanged. The most general solution of the equation of \ref{eq:max_lapl_fokker_planck} is 
\begin{equation} \label{eq:const_sol}
\tilde{Q_0}(x,s|M,m) = A e^{x \sqrt{s/D}} + B e^{-x \sqrt{s/D}} + \frac{1}{s}\,,
\end{equation}
where $A$ and $B$ are constants. Finally, imposing the boundary conditions, we obtain the solution
\begin{equation}\label{eq:span_survival}
    \tilde{Q_0}(x, s | M, m) =\frac1s\left[ 1 - \frac{\cosh(\sqrt{\frac{s}{4D}} (l - 2x))}{\cosh(\sqrt{\frac{s}{4D}} L)}\right],
\end{equation}
where we introduced two new variables
\begin{equation}\label{eq:Ll_def}
    L = M - m \mbox{~~and~~} l = M + m.
\end{equation}
The variable $L$ corresponds to the span of the RW, while $l$, which we denote \emph{imbalance}, corresponds to the maximum-minimum asymmetry of the RW. Indeed, trajectories with small $l$ will cover an almost symmetric (positive/negative) region, while for big values of $l$ the region will be highly asymmetric. Note that the joint distribution of the maximum $M$ and the minimum $m$ of a BM can also be derived using a path integral technique. This was done in Refs.~\cite{MMS19} and \cite{MMS20} where the distribution of the time between the maximum and the minimum was also computed.

We recall that $Q_0(x,t|M,m)$ corresponds to the cumulative distribution of the maximum $M$ and the minimum $m$ up to time $t$. Therefore, by differentiating Eq.~\ref{eq:span_survival} with respect to $M$ and $m$, and performing the change of variable $(M,m)\to(L=M-m,l=M+m)$, we find
\begin{equation}
\tilde{P}_0(L,l,s)=\int_{0}^{\infty}dt~e^{-st}P_0(L,l,t)=\frac{1}{4D}\frac{\cosh(l\sqrt{\frac{s}{4D}})}{\cosh^3(L\sqrt{\frac{s}{4D}})}\,,
\label{eq:P0Ll}
\end{equation}
where we have set $x=0$ and $P_0(L,l,t)$ denotes the joint PDF of the span $L$ and the imbalance $l$ at time $t$.

\begin{figure}
\centering
\includegraphics[width=0.5\textwidth]{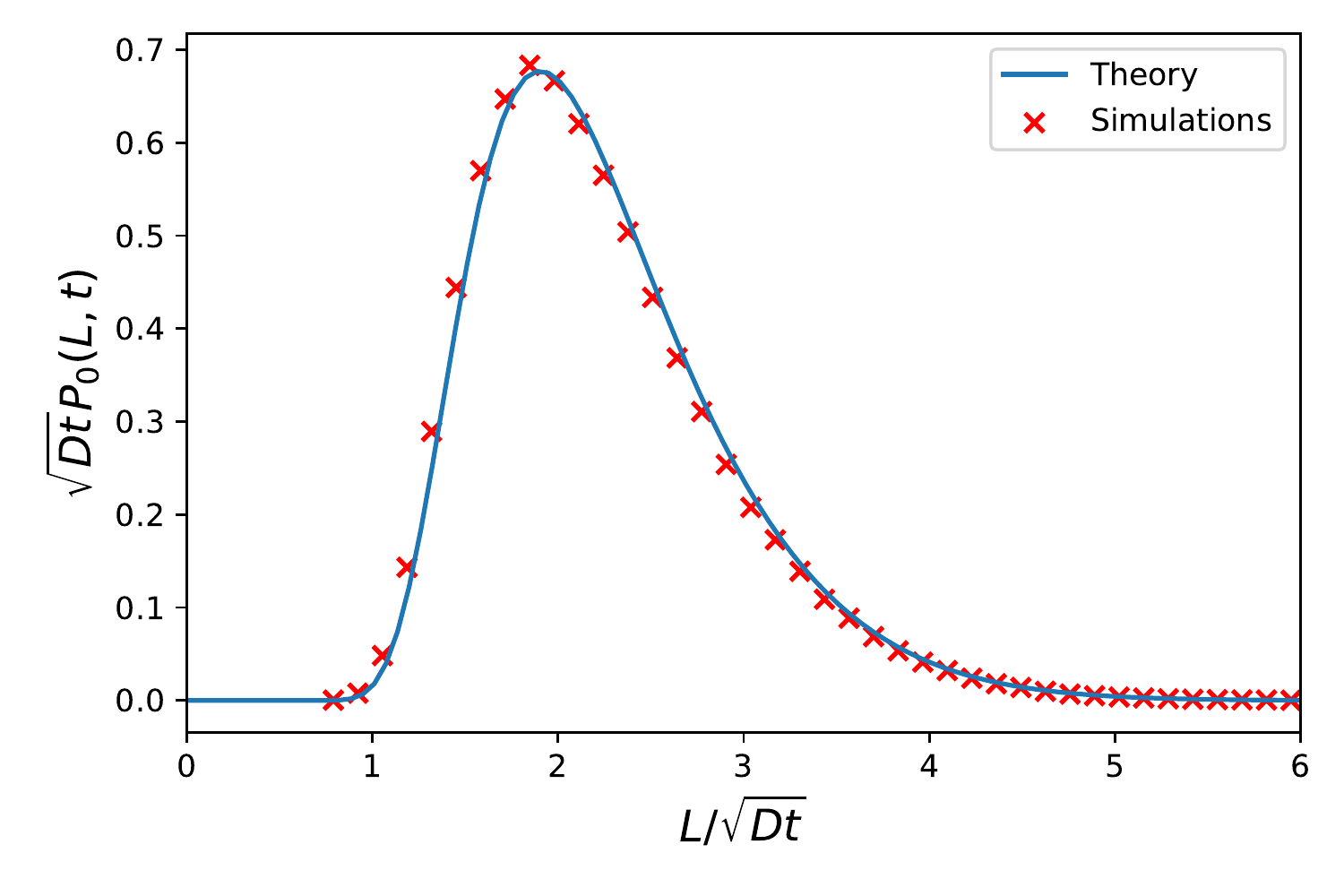}
\caption{The scaled distribution $\sqrt{Dt}P_0(L,t)$ as a function of the scaled span $L/\sqrt{Dt}$ for a Brownian motion of duration $t$ with diffusion coefficient $D$. The continuous blue line corresponds to the theoretical result in Eqs.~\ref{P0L} and \ref{G}. The symbols correspond to numerical simulations of Brownian motion performed with $dt=0.001$ and $D=1$.\label{fig:P0L}}
\end{figure}

Integrating this expression in Eq.~\ref{eq:P0Ll} over all possible values of $l$ with $L$ fixed, we obtain the marginal distribution of the span $L$
\begin{equation}
\tilde{P}_0(L,s)=\int_{-L}^{L}dl~\tilde{P}_0(L,l,s)=\frac{1}{\sqrt{sD}}\frac{\sinh(L\sqrt{\frac{s}{4D}})}{\cosh^3(L\sqrt{\frac{s}{4D}})}\,,
\end{equation}
Inverting the Laplace transform formally, we find
\begin{equation}
P_0(L,t)=\int_{\Gamma}\frac{ds}{2\pi i}e^{st}\frac{1}{\sqrt{sD}}\frac{\sinh(L\sqrt{\frac{s}{4D}})}{\cosh^3(L\sqrt{\frac{s}{4D}})}\,,
\label{eq:P0L}
\end{equation}
where the integral over $s$ runs over the imaginary-axis Bromwich contour. We identify the poles $s_k=-4D\pi^2(2k+1)^2/L^2$ of the integrand for $k\geq 0$ and we evaluate the residues at these poles to obtain
\begin{equation}
P_0(L,t)=\frac{1}{\sqrt{Dt}}\mathcal{G}\left(\frac{L}{\sqrt{Dt}}\right)\,,
\label{P0L}
\end{equation}
where
\begin{equation}
\mathcal{G}(z)=\frac{16}{z^3}\sum_{k=0}^{\infty}\left[\frac{2(2k+1)^2\pi^2}{z^2}-1\right]\exp\left(-\frac{\pi^2(2k+1)^2}{z^2}\right)\,,
\label{G}
\end{equation}
which coincides with the classical result of Feller \cite{F51}. This exact result is shown in Fig.~\ref{fig:P0L} and is in excellent agreement with numerical simulations. One can check that this scaling form is correctly normalized to unity and that it has asymptotic behaviors
\begin{equation}
\mathcal{G}(z)\approx\begin{cases}
32\pi^2e^{-\pi^2/z^2}/z^5 \,,&\text{ for }z\to 0\,,\\
\\
\frac{4}{\sqrt{\pi}}e^{-z^2/4}\,,&\text{ for }z\to \infty\,.
\end{cases}
\end{equation}
Note that to find the large-$z$ behavior we have used Poisson summation formula to obtain the alternative representation
\begin{equation}
\mathcal{G}(z)=\frac{4}{\sqrt{\pi}}\sum_{n=1}^{\infty}(-1)^{n+1}n^2  \exp(-\frac{n^2 z^2}{4})\,.
\end{equation}
This expression is also useful to compute the first moment of $L$, which reads
\begin{equation}
\langle L(t)\rangle= \frac{4\sqrt{Dt}}{\sqrt{\pi}}\,,
\end{equation}
in agreement with the expression obtained for the average number of visited sites $\langle V_0(n)\rangle$, given in Eq.~\ref{V0_d1} for a one-dimensional discrete-time random walker. To compare the two results, we recall that $D=a^2/(2 \Delta t)$, $n=t/\Delta t$, and $L(t)=V_0(n)/a$. Similarly, we find that the second moment of $L$ is given by
\begin{equation}
\langle L(t)^2\rangle=8\log(2) Dt\,.
\label{avg_L2}
\end{equation}

\begin{figure}
\centering
\includegraphics[width=0.5\textwidth]{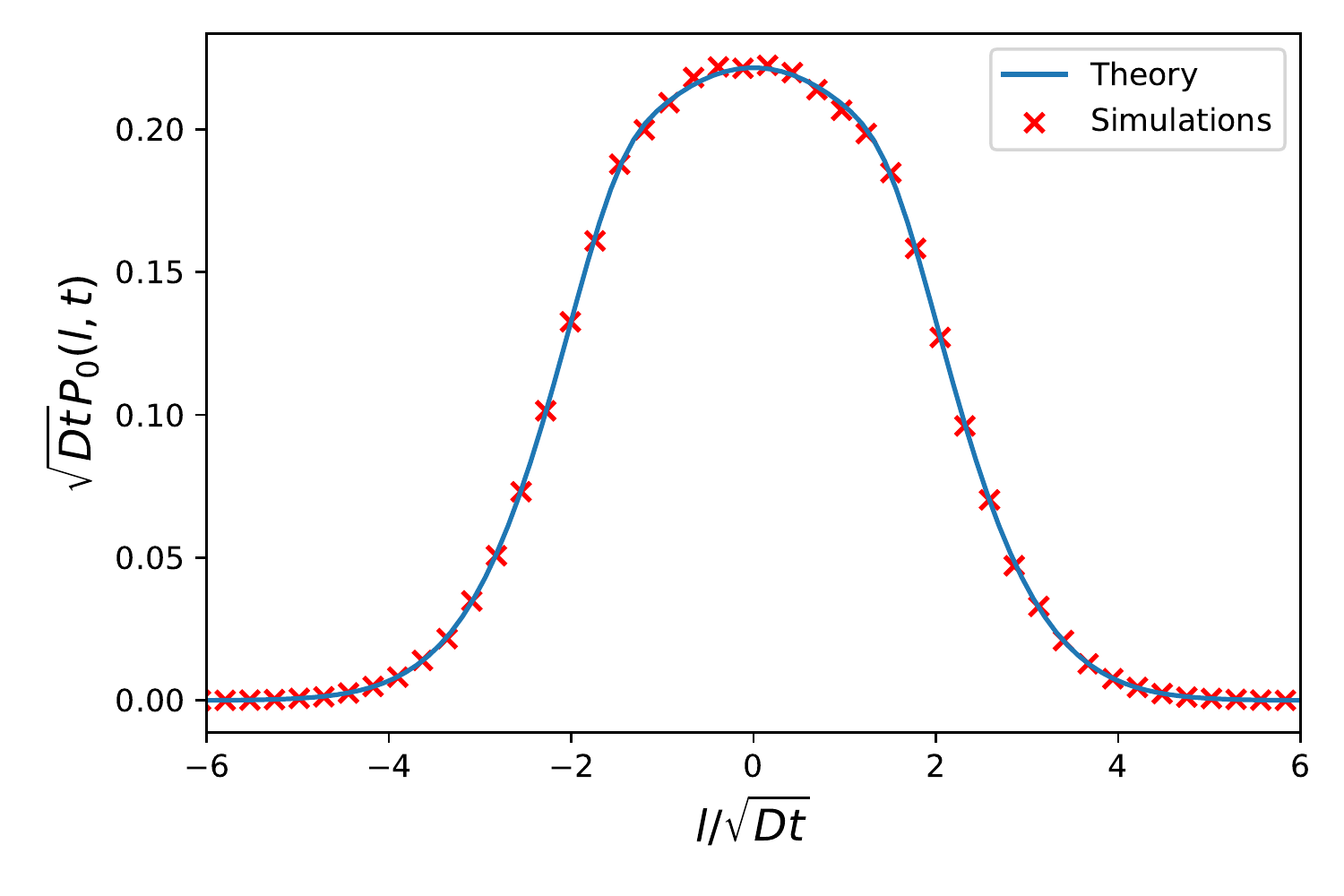}
\caption{The scaled distribution $\sqrt{Dt}P_0(l,t)$ as a function of the scaled imbalance $l/\sqrt{Dt}$ for Brownian motion of duration $t$ with diffusion coefficient $D$. The continuous blue line corresponds to the theoretical result in Eqs.~\ref{P_0l} and \ref{F}. The symbols correspond to numerical simulations performed with $dt=0.01$ and $D=1$.
\label{fig:P0l}}
\end{figure}

We now focus on the distribution of the imbalance $l$, which is defined as $l=M+m$. We recall a BM trajectory with a positive (negative) value of $l$ has mostly visited the region to the right (left) of the origin. Note that for a fixed value of $l$, the span $L$ can take values in $(|l|,\infty)$. Thus, integrating both sides of Eq.~\ref{eq:P0Ll} over $L$, we obtain
\begin{equation}
\tilde{P}_0(l,s)=\frac{\cosh(l\sqrt{\frac{s}{4D}})}{4D}\int_{|l|}^{\infty}dL~\frac{1}{\cosh^3(L\sqrt{\frac{s}{4D}})}\,,
\label{eq:P0l}
\end{equation}
Inverting the Laplace transform formally, we find
\begin{equation}
P_0(l,t)=\sqrt{Dt}\int_{\Gamma}\frac{ds}{2\pi i}~e^{st}\frac{\cosh(l\sqrt{\frac{s}{4D}})}{4D}\int_{|l|/\sqrt{Dt}}^{\infty}dy~\frac{1}{\cosh^3(y\sqrt{s t/4})}\,,
\label{eq:P0l2}
\end{equation}
where we have performed the change of variable $L\to y= L/\sqrt{Dt}$. As before, the integral over $s$ can be computed by identifying the poles on the right-hand side of Eq.~\ref{eq:P0l2} and evaluating the residues at these poles, yielding
\begin{equation}
P_0(l,t)=\frac{1}{\sqrt{Dt}}\mathcal{F}\left(\frac{l}{\sqrt{Dt}}\right)\,,
\label{P_0l}
\end{equation}
where 
\begin{align}
\mathcal{F}(z)\!  &= \!\frac{1}{2}\int_{|z|}^{\infty}dy\frac{1}{y^6}\sum_{k=0}^{\infty}(-1)^k e^{-\pi^2(2k+1)^2/y^2}\left\{(1+2k)\pi\left[y^2(24-y^2+z^2)-16(2k+1)^2\pi^2\right]\right.\nonumber\\
&\times \left. \cos(\frac{\pi(2k+1)z}{2y})+4zy\left[y^2-2(2k+1)^2\pi^2\right]\sin(\frac{\pi(2k+1)z}{2y})\right\}\,.
\label{F}
\end{align}
Note that, as a consequence of the $x\to-x$ symmetry of BM, $P_0(l,t)$ is symmetric around $l=0$ and the first moment vanishes
\begin{equation}
\langle l(t)\rangle=0\,.
\end{equation}
The second moment can be computed from the expression in Eq.~\ref{eq:P0l} and one obtains
\begin{equation}
\langle l(t)^2\rangle=8\log(\frac{e}{2})Dt\,.
\label{avg_l2}
\end{equation}
The asymptotic behaviors of the scaling function $\mathcal{F}(z)$ are given in Eq.~\ref{F_z_asymp}. The exact result in Eq.~\ref{P_0l} is shown in Fig.~\ref{fig:P0l} and is in perfect agreement with numerical simulation. We observe that the PDF of $l$ is maximal for $l=0$ and it decreases exponentially for large $l$.

Interestingly, from the expressions for the second moments of $L(t)$ and $l(t)$, respectively given in Eqs.~\ref{avg_L2} and \ref{avg_l2}, one can extract the covariance function 
\begin{equation}
\operatorname{cov}(m,M)=\langle M (t)m(t)\rangle-\langle M(t)\rangle\langle m(t)\rangle\,,
\end{equation}
which quantifies the correlations between the maximum $M(t)$ and the minimum $m(t)$ (recall that, by definition, $m(t)\!<\!0$). Indeed, since $L(t)\!=\!M(t)-m(t)$ and $l(t)\!=\!M(t)+m(t)$, we get
\begin{equation}
\langle M(t)m(t)\rangle=\frac14 \left[l(t)^2-L(t)^2\right]=-2\log(\frac{4}{e})Dt\,,
\end{equation}
where we have used Eqs.~\ref{avg_L2} and \ref{avg_l2}. Combining this expression with the result $\langle M(t)\rangle=-\langle m(t)\rangle=(2/\sqrt{\pi})\sqrt{Dt}$, we obtain
\begin{equation}
\operatorname{cov}(m,M)=\left[\frac{4}{\pi}-2\log(\frac{4}{e})\right]Dt\approx 1.48417~ Dt\,,
\end{equation}
meaning that the maximum and the minimum are positively correlated. Note that since $m(t)<0$ this implies that the magnitudes of the maximum and the minimum are actually strongly anticorrelated.

\subsection{The case $r > 0$}

We next focus on the case of resetting $r>0$. To proceed, we define the survival probability $Q_r(x,t|M,m)$ as the probability that a Brownian motion with resetting rate $r>0$ and with initial position $x$ always remains inside the interval $[m,M]$ up to time $t$. As for the case without resetting, this quantity is the joint cumulative distribution of the maximum $M$ and the minimum $m$. In the presence of resetting, the survival probability satisfies the renewal equation
\begin{equation}
Q_r(x, t | M, m) = e^{-r t} Q_0(x, t | M, m) + r \int_0^t Q_r(x, t - \tau | M, m) Q_0(x, \tau | M, m)e^{-r\tau} \dd \tau\,.\label{renewal_Qr}
\end{equation}
The first term on the right-hand side corresponds to the case where no resetting occurs up to time $t$, where $e^{-rt}$ is the corresponding probability. In this case, the survival probability is $Q_0(x, t | M, m)$, i.e., the one without resetting. The second term in Eq.~\ref{renewal_Qr} corresponds instead to the case where the first resetting event occurs at time $\tau$, with probability weight $re^{-r\tau}$. Taking the Laplace transform of the above equation and setting $x = 0$, we obtain
\begin{equation}\label{eq:last_reset_lapl}
\tilde{Q}_r(0, s | M, m) = \frac{\tilde{Q}_0(0, s + r | M , m)}{1 - r \tilde{Q}_0(0, s + r | M , m)}\,,
\end{equation}
where $\tilde{Q}_r(x, s | M, m)$ is the Laplace transform of $Q_r(0, t | M, m)$ with respect to $t$. Plugging the expression for $\tilde{Q}_0(0, s | M, m)$ given in Eq.~\ref{eq:span_survival} into Eq.~\ref{eq:last_reset_lapl}, we find
\begin{equation} \label{eq:cdf_Ll_reset}
\tilde{Q_r}(x = 0, s | L, l) = \frac{\cosh(L\sqrt{\frac{s + r}{4 D}}) - \cosh(l \sqrt{\frac{s + r}{4 D}})}{r \cosh(l \sqrt{\frac{s + r}{4 D}}) + s \cosh(L \sqrt{\frac{s + r}{4 D}})}.
\end{equation}
Differentiating with respect to $M$ and $m$ and performing the change of variables $(M, m) \to (L = M - m,\, l = M + m)$ we find that the joint distribution of $L$ and $l$ reads
\begin{equation}
\tilde{P_r}(L, l, s) = \frac{(r+s)^2}{4 D} \frac{r \cosh(L\sqrt{\frac{s + r}{4D}}) + s \cosh(l \sqrt{\frac{s + r}{4D}})}{\left( s \cosh(L\sqrt{\frac{s + r}{4D}}) + r \cosh(l\sqrt{\frac{s + r}{4D}}) \right)^3}.
\end{equation}
This equation cannot be simply reduced to a scaling form. However, we can make progress in the long-time limit, corresponding to $s\to 0$, where
\begin{equation}\label{eq:pdf_Ll_small_s_lapl}
\tilde{P_r}(L, l, s)\approx \frac{r^3}{4 D}  \frac{\sech^2(L\sqrt{\frac{r}{ 4D}})}{\left( s + r \cosh(l\sqrt{\frac{r}{4D}}) \sech(L \sqrt{\frac{r}{4D}}) \right)^3}.
\end{equation}
The Laplace transform can now be inverted, yielding
\begin{equation} \label{eq:pdf_Ll_small_s}
P_r(L, l, t) \approx\frac{r^3 t^2}{8 D}\frac{\exp( -r t \cosh(l \alpha_0/2)\sech(L \alpha_0/2) )}{\cosh^2(L \alpha_0/2)},
\end{equation}
where we introduced the constant $\alpha_0 = \sqrt{r/D}$. 

\begin{figure}
\centering
\includegraphics[width=0.5\textwidth]{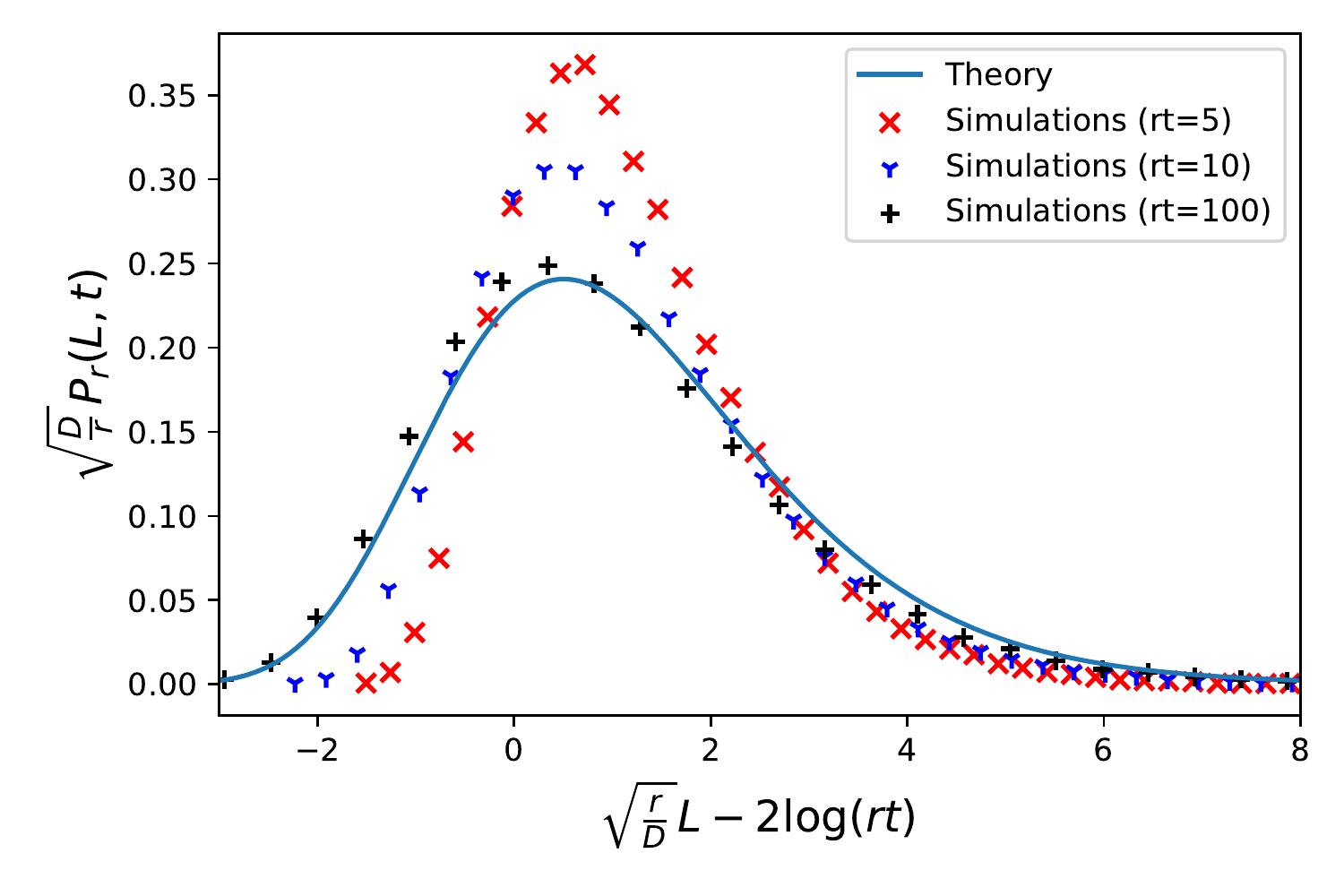}
\caption{The scaled distribution $\sqrt{D/r}P_r(L,t)$ as a function of the scaled span $\sqrt{r/D}L-2\log(rt)$ for resetting Brownian motion of duration $t$ with diffusion coefficient $D$ and resetting rate $r$. The continuous blue line corresponds to the theoretical result in Eqs.~\ref{PL_scale} and \ref{g}, valid for $t\gg1/r$. The symbols correspond to numerical simulations performed with $dt=0.01$, $D=1$, and different values of $rt$. As $t$ increases, the results of numerical simulations approach the scaling form in Eqs.~\ref{PL_scale} and \ref{g}.\label{fig:1d_pdf}}
\end{figure}

To get the marginal PDF of the span $L$, we integrate this PDF in Eq.~\ref{eq:pdf_Ll_small_s} over $l \in [-L, L]$, which gives
\begin{equation} 
P_r(L, t) \approx \frac{r^3 t^2}{4 D {\cosh^2(L \alpha_0/2)}} \int_{0}^{L} \exp[ -r t\cosh(l \alpha_0/2)\sech(L\alpha_0/2)] \dd l.
\end{equation}
In the long time limit, we can expand for $L\gg 1$ and we obtain
\begin{equation} 
P_r(L, t) \approx \frac{r^3 t^2}{4 D {\cosh^2(L \alpha_0/2)}} \int_{0}^{+\infty} \exp( -2r t \cosh(l \alpha_0/2)e^{-\alpha_0 L/2}) \dd l.
\end{equation}
Computing the integral over $l$ \cite{Bessel}, we obtain the scaling form
\begin{equation}
P_r(L, t) \approx  \alpha_0 ~g\left[\alpha_0(L-2\alpha_0^{-1}\log(rt))\right],
\label{PL_scale}
\end{equation}
where 
\begin{equation}
g(z) = 2e^{-z} K_0\left(2e^{-z/2}\right),
\label{g}
\end{equation}
Note that the variable $z$ takes values in $(-\infty,\infty)$ for large $rt$. The asymptotic behaviors of the scaling function are given in Eq.~\ref{g_asympt}. Comparison between the theoretical curve and the numerical simulation is displayed in Figure \ref{fig:1d_pdf} in which we can see a perfect agreement. Interestingly, this scaling function $g(z)$ also describes the distribution of the number of distinct sites visited by $N$ random walkers in the limit of large $N$ \cite{KMS13}.

It is easy to check that the function $g(z)$ is positive and normalized to unity \ref{app:g}. Thus, at late times, the span $L(t)$ can be written as
\begin{equation}
L(t)\approx 2\sqrt{\frac{D}{r}}\log(rt)+\sqrt{\frac{D}{r}}z\,,
\label{Lz}
\end{equation}
where $z$ is a random variable with PDF $g(z)$. In other words, at late times, the span $L(t)$ becomes deterministic to leading order, with random contribution of order one. The first moment of $L(t)$ can be obtained as
\begin{equation}
\langle L(t)\rangle \approx 2\sqrt{\frac{D}{r}}\log(rt)+\sqrt{\frac{D}{r}}\langle z\rangle \approx 2\sqrt{\frac{D}{r}}\left[\log(rt)+\gamma_E\right]\,,\label{eq:1st_moment_L}
\end{equation}
where we have used $\langle z\rangle=2\gamma_E$ \ref{app:g}. Notice that this corresponds to exactly twice of the average of the maximum which was computed in \cite{MajumdarConvex}. This is expected since
\begin{equation}
\langle L \rangle = \langle M - m \rangle = \langle M \rangle + \langle - m \rangle = 2 \langle M \rangle.\label{relationLM}
\end{equation}

Similarly, using Eq.~\ref{Lz}, we can write the second moment of $L(t)$ at late times as
\begin{equation}
\langle L(t)^2\rangle\approx 4\frac{D}{r}\log^2(rt)+\frac{D}{r}\langle z^2\rangle+2\frac{D}{r}\log(rt)\langle z\rangle\,.
\end{equation}
Using the expressions for $\langle z\rangle$ and $\langle z^2\rangle$ \ref{app:g}, we obtain
\begin{equation}
    \langle L^2 \rangle \approx 4\alpha_0^{-2} \left( (\log(rt) + \gamma_E)^2 + \frac{\pi^2}{12} \right) = 4 \frac{D}{r} \left( (\log(rt) + \gamma_E)^2 + \frac{\pi^2}{12} \right) \label{eq:2nd_moment_L}
\end{equation}
Hence putting equation \ref{eq:1st_moment_L} and \ref{eq:2nd_moment_L} together we can obtain the variance of the span which is given by, for $t\gg 1/r$,
\begin{equation}\label{eq:var_L}
\text{Var}(L) = \langle L^2 \rangle - \langle L \rangle^2 \approx \frac{D}{r} \frac{\pi^2}{3}\,.
\end{equation}

\begin{figure}
\centering
\includegraphics[width=0.5\textwidth]{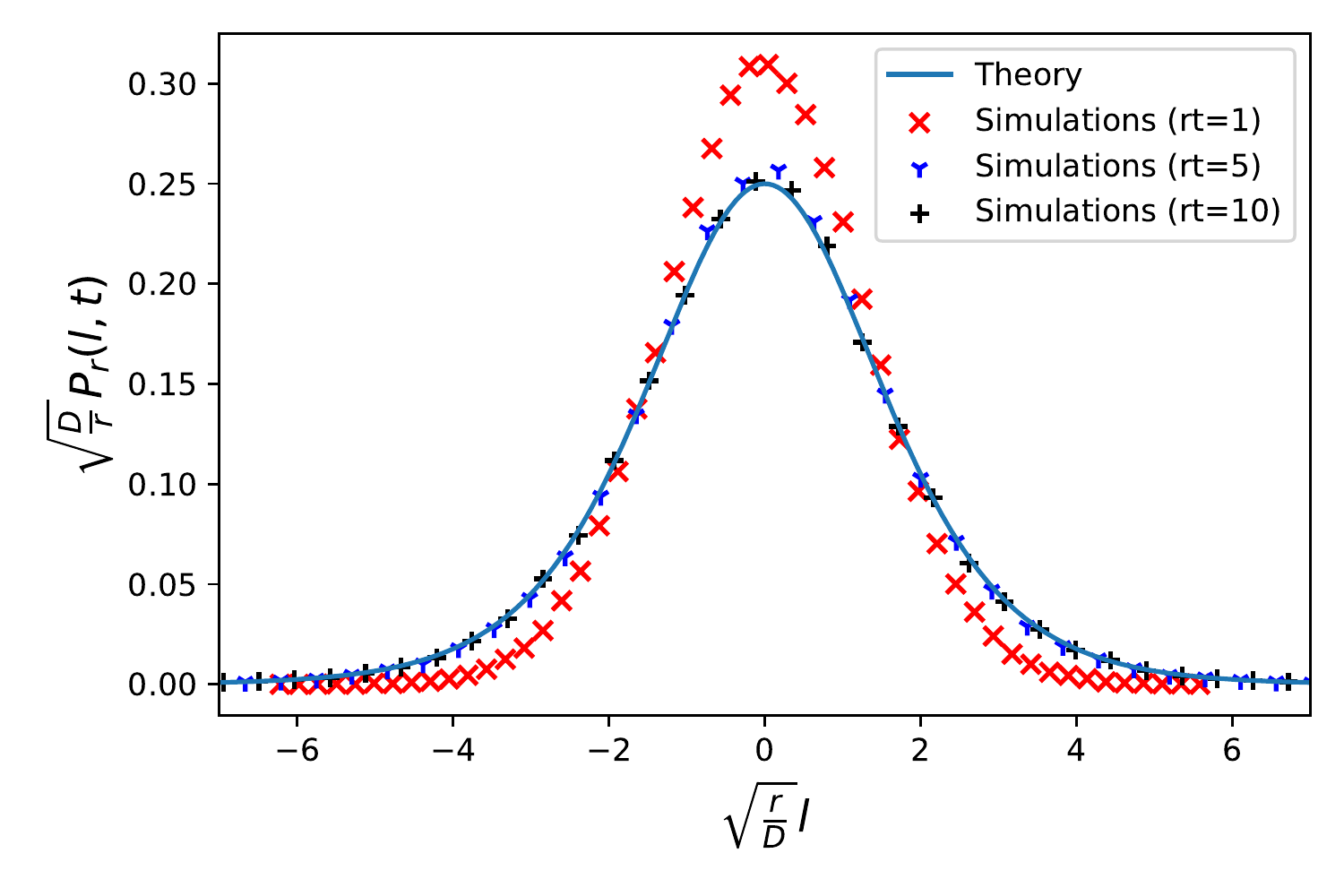}
\caption{The scaled distribution $\sqrt{D/r}P_r(l,t)$ as a function of the scaled imbalance $\sqrt{r/D}l$ for resetting Brownian motion of duration $t$ with resetting $r$ and diffusion coefficient $D$. The continuous blue line corresponds to the theoretical result in Eqs.~\ref{Prlt_1} and \ref{f}, valid for $t\gg1/r$. The symbols correspond to numerical simulations performed with $dt=0.01$, $D=1$ and different values of $rt$. As $t$ increases, the results of numerical simulations approach the scaling form in Eqs.~\ref{Prlt_1} and \ref{f}.\label{fig:1d_l_pdf}}
\end{figure}

We next focus on the late-time distribution of the imbalance $l$. Integrating over $L$ the joint distribution of $L$ and $l$, given in Eq.~\ref{eq:pdf_Ll_small_s}, we obtain
\begin{equation} 
P_r( l, t) \approx\frac{r^3 t^2}{8 D}\int_{|l|}^{\infty}dL~\frac{\exp( -r t \cosh(l \alpha_0/2)\sech(L \alpha_0/2) )}{\cosh^2(L \alpha_0/2)}\,.
\end{equation}
For $t\gg 1/r$, the integral will be dominated by large values of $L$, yielding
\begin{equation} 
P_r(L, l, t) \approx\frac{r^3 t^2}{2 D}\int_{0}^{\infty}dL~\exp( -2r t \cosh(l \alpha_0/2) e^{-\alpha_0 L/2})e^{-\alpha_0 L}\,.
\end{equation}
Performing the integral over $L$, we find
\begin{equation}
P_r(l,t)\approx \sqrt{\frac{r}{D}}h\left(\sqrt{\frac{r}{D}}\,\,l\right)
\label{Prlt_1}
\end{equation}
where
\begin{equation}
h(y)=\frac{1}{4\cosh^2(y/2)}\,.
\label{f}
\end{equation}
The distribution $h(y)$ is known in the probability theory literature as logistic distribution. The scaling function in Eqs.~\ref{Prlt_1} and \ref{f} are shown in Fig. \ref{fig:1d_l_pdf} and are in perfect agreement with numerical simulations. It is easy to check that $h(y)$ is correctly normalized to unity in $(-\infty,\infty)$. Its asymptotic behaviors are given in Eq.~\ref{h_asymp}. Moreover, since $h(y)$ is symmetric around $y=0$, the first moment $\langle y\rangle$ vanishes. On the other hand, one can check that the second moment is given by
\begin{equation}
\langle y^2\rangle =\frac{\pi^2}{3}\,.
\end{equation}
Thus, the first and second moments of $l(t)$ are given by
\begin{equation}
\langle l \rangle = 0 \mbox{~~and~~} \text{Var}(l) \approx \frac{D}{r} \frac{\pi^2}{3}.
\end{equation}
In summary, we have shown that, even though the number of visited sites grows in time, the fractions of these sites that are on the positive and negative sides will be roughly the same, up to an order-one random correction.

The late-time distribution of $L(t)$ can also be computed using the following simple argument. We recall that the span is defined as $L(t)=M(t)-m(t)$, where $M(t)$ and $m(t)$ are the maximum and the minimum of the process. For $t\gg 1/r$, we expect that several resetting events have already occurred and the time interval $[0,t]$ can be split into independent time intervals, separated by the resetting events. For late times, it is reasonable to expect that the minimum $m(t)$ and the maximum $M(t)$ are not reached in the same interval. This implies that $M(t)$ and $m(t)$ become independent for large $t$. The distribution of the maximum $M(t)$ tends at late times to a Gumbel distribution. In particular, in \cite{Evans}, it was shown that, for $t\gg 1/r$,
\begin{equation}
M(t)\approx \sqrt{\frac{D}{r}}\log(rt)+ \sqrt{\frac{D}{r}} u_1\,,
\label{Mlt}
\end{equation}
where $-\infty<u_1<\infty$ is a Gumbel random variable with probability density function
\begin{equation}
P(u)=e^{-u-e^{-u}}\,.
\label{Pu}
\end{equation}
Similarly, due to the $x\to -x$ symmetry of the process, the minimum $m(t)$ goes at late times as
\begin{equation}
m(t)\approx-\sqrt{\frac{D}{r}}\log(rt)- \sqrt{\frac{D}{r}} u_2\,,
\label{mlt}
\end{equation}
where $u_2$ is drawn from $P(u)$. Moreover, at late times, we expect $u_1$ and $u_2$ to become uncorrelated. Thus, using the definition of $L(t)$, \ref{Mlt}, and \ref{mlt}, we obtain
\begin{equation}
L(t)\approx 2\sqrt{\frac{D}{r}}\log(rt)+ \sqrt{\frac{D}{r}} z\,,
\end{equation}
where $z=u_1+u_2$ is the sum of two i.i.d. Gumbel variables. The distribution of $z$ is then simply given by
\begin{equation}
g(z)=\int_{-\infty}^{\infty}du~P(u)P(z-u)\,.
\end{equation}
Plugging the expression of $P(u)$, given in \ref{Pu}, we obtain \cite{Bessel}
\begin{equation}
g(z)=e^{-z}\int_{-\infty}^{\infty}du~\exp\left[-e^{-u}-e^{-(z-u)}\right]=2^{-z}K_0(2 e^{-z/2})\,,
\end{equation}
in agreement with \ref{g}.

With a similar argument one can also compute the late-time distribution of the imbalance $l(t)$, defined as the sum between the maximum and the minimum of the process up to time $t$. Using \ref{Mlt} and \ref{mlt}, we obtain
\begin{equation}
 l(t)=M(t)+m(t)=\sqrt{\frac{D}{r}}y\,,
\end{equation}
with $y=u_1-u_2$, where $u_1$ and $u_2$ are independently drawn from $P(u)$ in \ref{Pu}. Thus, the late-time PDF of $l(t)$ is given by
\begin{equation}
P_r(l,t)\approx \sqrt{\frac{r}{D}}h\left(\sqrt{\frac{r}{D}}\,\,l\right)\,,
\label{Pl_scal}
\end{equation}
where $h(y)$ is the PDF of $y$, which can be written as
\begin{equation}
h(y)=\int_{-\infty}^{\infty}\dd u~P(u)P(y+u)\,.
\end{equation}
Using Eq.~\ref{Pu}, we recover the logistic distribution in Eq.~\ref{f}. Indeed, it is well-known that the difference between two independent Gumbel random variables has a logistic distribution.

Note that even though here we have focused on the late-time limit, corresponding to $t\gg1/r$, it is possible to compute exactly the average span $\langle L(t)\rangle$ for any finite $t$. Indeed, the full time dependence of $\langle M(t)\rangle$ was recently derived in \cite{MajumdarConvex}. Combining the result from \cite{MajumdarConvex} with Eq.~\ref{relationLM}, we obtain
\begin{equation}
\langle L(t)\rangle = 2\sqrt{\frac{D}{r}}F_1(rt)\,,
\end{equation}
where
\begin{equation}
F_1(z)=\int_{0}^{z}dy~\frac{1}{y}(1-e^{-y})\left[\frac{e^{-(z-y)}}{\sqrt{\pi(z-y)}}+\operatorname{erf}(\sqrt{z-y})\right]\,.
\end{equation}
This result coincides with the one derived in Section \ref{sec:scaling} for the scaling regime in $d=1$, which indeed corresponds to the limit of resetting BM. As a final remark, it is interesting to notice that the function $F_1(z)$ appears also in the context of record statistics of a discrete-time continuous-space RW \cite{MMS21}. Indeed, in Ref. \cite{MMS21}, it was shown that the average number of records $\langle R_n\rangle$ of a resetting RW of $n$ steps grows at late times as
\begin{equation}
\langle R_n\rangle\approx\frac{1}{\sqrt{p}}F_1(pn)\,,
\end{equation}
where $p$ is the resetting probability.

\section{Conclusion}

\label{sec:conclusion}

In this paper, we have computed the average number of visited sites $\langle V_p(n) \rangle$ for an $n$-step resetting RW with resetting probability $p$ in the limit of large $n$. We have shown that on average the number of visited sites grows very slowly as $[\log(n)]^d$ for any $d>0$. At variance with the case of random walks without resetting, we have shown that the recurrence-transience transition at $d=2$ for standard random walks (with no resetting) disappears in the presence of resetting. In the limit $p\to 0$ and for any $d>0$, we have derived the exact expression of the crossover scaling function of $\langle V_p(n) \rangle$. This function interpolates between the regimes $p=0$ and $p>0$. In $d=2$, our results are in agreement with those of Ref.~\cite{MajumdarConvex}, where the average area of the convex hull of a two-dimensional resetting BM was computed. In particular, we have observed that in $d=2$ the average number of visited sites coincides to leading order with the average area of the convex hull, indicating that almost every site in the convex hull is visited for large $n$. 

For $d=1$, we have considered the continuum limit where the resetting random walker can be approximated as a resetting Brownian motion. In this limit, we have derived the full distribution of the number of visited sites for late times. Moreover, we have introduced a new observable, the imbalance, which quantifies how much the visited region is symmetric around the starting position in $d=1$. We have computed the exact distribution of the imbalance both with and without resetting. As a result, we have shown that the region visited by a resetting random walker becomes symmetric around the resetting location for late times.

Following the same principles used in this paper to obtain the exact expression of the average number of visited sites, it is also possible to obtain an exact expression for the second moment. However, deriving the long-time asymptotic behavior is considerably more complicated. It would be interesting to be able to derive the asymptotics of the second moment in order to have an expression for the variance. Moreover, even though in this work we have focused on the behavior of a resetting random walk on a hypercubic lattice, it would be interesting to consider more complicated graph topologies as well as resetting to multiple locations. For arbitrary regular lattices in $d$-dimensions, we expect the late-times asymptotic behavior of $\langle V_p(n)\rangle$ in Eq.~\ref{eq:Vpn_intro} to be universal, i.e., independent of the specific details of the lattice. However, this result will not be valid for more complicated topologies, e.g., for trees. Finally, it would be also interesting to compute the distribution of $V_p(n)$ in $d>1$.

\appendix
\section{Derivation of Eq.~\ref{eq:G0_expression}}

\label{app:rec}

In this appendix, we derive the expression for the propagator $G_0(\vb{0},\vb{X},n)$, given in Eq.~\ref{eq:G0_expression}. Our starting point is the recursion relation
\begin{equation}
G_0(\vb{0},\vb{X},n)=\frac{1}{2d}\sum_{i=1}^{d}\left[G_0(\vb{0},\vb{X}+\vb{e}_i,n-1)+G_0(\vb{0},\vb{X}-\vb{e}_i,n-1)\right]\,.
\label{rec_app}
\end{equation}
Taking a discrete Fourier transform on both sides of Eq.~\ref{rec_app}, we obtain
\begin{equation}
\tilde{G}_0(\vb{k},n)=\tilde{G}_0(\vb{k},n-1) \frac{1}{d}\sum_{i=1}^d \cos(k_i)\,,\label{rec2_app}
\end{equation}
where we have defined
\begin{equation}
\tilde{G}_0(\vb{k},n)=\sum_{\vb{X}}e^{i\vb{k}\cdot \vb{X}}G_0(\vb{0},\vb{X},n)\,,
\end{equation}
and the sum over $\vb{X}$ runs over all integers in $\Z^d$. Repeatedly applying Eq.~\ref{rec2_app}, we obtain
\begin{equation}
\tilde{G}_0(\vb{k},n)=\tilde{G}_0(\vb{k},0)\left[ \frac{1}{d}\sum_{i=1}^d cos(k_i)\right]^n\,.\label{rec3_app}
\end{equation}
From the initial condition $G_0(\vb{0},\vb{X},0)=\delta_{\vb{X},0}$ we obtain $\tilde{G}_0(\vb{k},0)=1$, yielding the result in Eq.~\ref{eq:G0_expression}.

\section{Asymptotic behaviors of $\tilde{G}_0(\vb{0},\vb{0},z)$}
\label{app:G_lim}

In this appendix, we investigate the asymptotic behavior of $\tilde{G}_0(\vb{0},\vb{0},1-s)$ in the limit of small $s$. We use the exact expression
\begin{equation}
\tilde{G}_0(\vb{0},\vb{0},1-s)=\int_{-\pi}^{\pi}\ldots\int_{-\pi}^{\pi}\frac{d^d\vb{k}}{(2\pi)^d}\frac{1}{1-\frac{1-s}{d}\sum_{i=1}^{d}\cos k_i}\,.
\end{equation}

For $d<2$, the integral on the right hand side is dominated by small values of $k_i$ when $s\to 0$, yielding
\begin{equation}
\tilde{G}_0(\vb{0},\vb{0},1-s)\approx\int_{-\pi}^{\pi}\ldots\int_{-\pi}^{\pi}\frac{d^d\vb{k}}{(2\pi)^d}\frac{1}{s+k^2/(2d)}\,,
\label{G0App}
\end{equation}
where $k^2=\sum_i k_i^2$. Performing the change of variable $k_i\to q_i=k_i/\sqrt{s}$, we find
\begin{equation}
\tilde{G}_0(\vb{0},\vb{0},1-s)\approx s^{d/2-1} \int_{-\infty}^{\infty}\ldots\int_{-\infty}^{\infty}\frac{d^d\vb{q}}{(2\pi)^d}\frac{1}{1+q^2/(2d)}=A_ds^{d/2-1}\,,
\end{equation}
where $A_d$ is given in Eq.~\ref{A_d}.

Similarly, it is possible to show that the integral over $\vb{k}$ converges in the limit $s\to0$ for $d>2$. One therefore obtains
\begin{equation}
\tilde{G}_0(\vb{0},\vb{0},1-s)\approx B_d=\int_{-\pi}^{\pi}\ldots\int_{-\pi}^{\pi}\frac{d^d\vb{k}}{(2\pi)^d}\frac{1}{1-\frac{1}{d}\sum_{i=1}^{d}\cos k_i}\,,
\end{equation}
with $B_d<\infty$.

Finally, we consider the case $d=2$. Also in this case, the integral in Eq.~\ref{G0App} is dominated by small values of $k_i$ when $s\to 0$, yielding
\begin{equation}
\tilde{G}_0(\vb{0},\vb{0},1-s)\approx\frac{1}{4\pi^2}\int_{-\pi}^{\pi}dk_1~\int_{-\pi}^{\pi}dk_2~\frac{1}{s+k^2/(2d)}\,.
\end{equation}
Performing the change of variable $k_i\to q_i=k_i/\sqrt{s}$, we obtain
\begin{equation}
\tilde{G}_0(\vb{0},\vb{0},1-s)\approx\frac{1}{4\pi^2}\int_{-\pi/\sqrt{s}}^{\pi/\sqrt{s}}dq_1~\int_{-\pi/\sqrt{s}}^{\pi/\sqrt{s}}dq_2~\frac{1}{1+q^2/(2d)}\,.
\end{equation}
Passing to spherical coordinates, we find
\begin{equation}
\tilde{G}_0(\vb{0},\vb{0},1-s)\approx\frac{1}{2\pi}\int_{0}^{\pi/\sqrt{s}}dq~\frac{q}{1+q^2/(2d)}=\frac{1}{\pi}\log(\frac1s)\,.
\end{equation}

\section{Laplace inversion of Eqs.~\ref{d2_LT} and \ref{identity_Vp6}}
\label{app:laplace_inv}
In this appendix we want to show that to leading order for small $s$
\begin{equation}
\int_{0}^{\infty}\dd t ~\frac{t}{\log(t)}e^{-st}\approx \frac{1}{s^2 \log(1/s)}\,.
\label{eq:lt_app}
\end{equation}
Performing the change of variable $t\to q=st$, we obtain
\begin{equation}
\int_{0}^{\infty}\dd t ~\frac{t}{\log(t)}e^{-st}=\frac{1}{s^2}\int_{0}^{\infty}\dd q ~\frac{q}{\log(q)+\log(1/s)}e^{-q}\,.
\end{equation}
For small $s$, we can neglect the term $\log(q)$ on the right-hand side, yielding the result in Eq.~\ref{eq:lt_app}.

To invert the Laplace transform in Eq.~\ref{identity_Vp6}, we show that at leading order for small $s$
\begin{equation}
\int_{0}^{\infty}dt~\left[\log(t)\right]^de^{-st}\approx \frac1s \left[\log(\frac{1}{s})\right]^d\,.
\end{equation}
Indeed, performing the change of variable $t\to q=st$, we obtain
\begin{equation}
\int_{0}^{\infty}dt~\left[\log(t)\right]^de^{-st}=\frac1s \int_{0}^{\infty}dq~\left[\log(q)+\log(\frac1s)\right]^de^{-q}\,.
\end{equation}
Thus, to leading order for small $s$, we find
\begin{equation}
\int_{0}^{\infty}dt~\left[\log(t)\right]^de^{-st}\approx\frac1s \left[\log(\frac1s)\right]^d \int_{0}^{\infty}dq~e^{-q}=\frac1s \left[\log(\frac1s)\right]^d \,.
\end{equation}

\section{Asymptotic behavior of $\tilde{G}_0(\vb{0},\vb{X},1-p)$ for large $|\vb{X}|$}

\label{app:asympt_G0}
In this appendix, we extract the large-$|\vb{X}|$ asymptotic behavior of $\tilde{G}_0(\vb{0},\vb{X},1-p)$. We start from the exact expression (see Eq.~\ref{eq:Gz0_expression})
\begin{equation}
\tilde{G}_0(\vb{0},\vb{X},1-p)=\int_{-\pi}^{\pi}\ldots\int_{-\pi}^{\pi}\frac{d^d\vb{k}}{(2\pi)^d}e^{-i \vb{k}\cdot\vb{X}}\frac{1}{1-((1-p)/d)\sum_{i=1}^{d}\cos k_i}\,.
\end{equation}
For large $|\vb{X}|$, the integral on the right-hand side is dominated by small values of $\vb{k}$. Expanding for small $k_i$, we obtain 
\begin{equation}
\tilde{G}_0(\vb{0},\vb{X},1-p)\approx \frac{2d}{1-p}\int_{-\pi}^{\pi}\ldots\int_{-\pi}^{\pi}\frac{d^d\vb{k}}{(2\pi)^d}e^{-i \vb{k}\cdot\vb{X}}\frac{1}{k^2 +2dp/(1-p)}\,,
\label{G0_smallk}
\end{equation}
where $k^2=k_1^2 +\ldots k_d^2$. Using the identity
\begin{equation}
\frac1a=\int_{0}^{\infty}dt~e^{-at}\,,
\end{equation}
we rewrite Eq.~\ref{G0_smallk} as
\begin{equation}
\tilde{G}_0(\vb{0},\vb{X},1-p)\approx \frac{2d}{1-p}\int_{-\pi}^{\pi}\ldots\int_{-\pi}^{\pi}\frac{d^d\vb{k}}{(2\pi)^d}e^{-i \vb{k}\cdot\vb{X}}\int_{0}^{\infty}dt~e^{-(k^2 +2dp/(1-p))t}\,.
\label{G0_smallk2}
\end{equation}
We can now perform the Gaussian integrals over $k_i$, yielding
\begin{equation}
\tilde{G}_0(\vb{0},\vb{X},1-p)\approx \frac{2d}{1-p}\int_{0}^{\infty}dt~\frac{1}{(4\pi t)^{d/2}}e^{-2dp t/(1-p)-|\vb{X}|^2/(4t)}\,.
\end{equation}
Performing the integral over $t$, we find
\begin{equation}
\tilde{G}_0(\vb{0},\vb{X},1-p)\approx \frac{1}{2p (4\pi)^{d/2}}\left(\frac{1-p}{8dp}\right)^{-(d+2)/4}|\vb{X}|^{1-d/2}K_{1-d/2}\left(\sqrt{\frac{2dp}{1-p}}|\vb{X}|\right)\,,
\end{equation}
where $K_{\nu}(z)$ is the modified Bessel function of the second kind and we have use the identity \cite{Bessel}
\begin{equation}
\int_{0}^{\infty}dt~t^{\nu-1}e^{-\beta /t-\gamma t}=2\left(\frac{\beta}{\gamma}\right)^{\nu/2}K_{\nu}(\sqrt{4\beta \gamma})\,.
\end{equation}

\section{Laplace inversion of Eq.~\ref{lapl_F1}}

\label{app:F_1_inversion}

In this appendix, we present the derivation of the Laplace inversion of Eq.~\ref{lapl_F1}, which can be done by convolution theorem. We denote 
\begin{equation}
g_1(y)=\int_{\Gamma}\frac{dq}{2\pi i}e^{qy}\log(\frac{q+1}{q})\,,
\end{equation}
\begin{equation}
g_2(y)=\int_{\Gamma}\frac{dq}{2\pi i}e^{qy}\frac{\sqrt{1+q}}{q}\,.
\label{g21}
\end{equation}
Then, by convolution theorem, we have
\begin{equation}
F_1(z)= \int_{0}^{\infty}dy~g_1(y)g_2(z-y)\,.
\label{convol_F1}
\end{equation}

The function $g_1(y)$ can be obtained using the identity
\begin{equation}
\int_{0}^{\infty}dy~e^{-qy}\frac{1-e^{-y}}{y}=\log(\frac{q+1}{q})\,,
\end{equation}
yielding 
\begin{equation}
g_1(y)=\frac{1-e^{-y}}{y}\,.
\end{equation}

To compute $g_2(y)$, we first rewrite Eq.~\ref{g21} as
\begin{equation}
g_2(y)=\int_{\Gamma}\frac{dq}{2\pi i}e^{qy}\frac{1}{\sqrt{1+q}}+\int_{\Gamma}\frac{dq}{2\pi i}e^{qy}\frac{1}{q\sqrt{1+q}}\,.
\end{equation}
The first term can be immediately computed and reads
\begin{equation}
\int_{\Gamma}\frac{dq}{2\pi i}e^{qy}\frac{1}{\sqrt{1+q}}=\frac{e^{-y}}{\sqrt{\pi y}}
\label{lapla_1_g1}
\end{equation}
The second term can be inverted using the relation in Eq.~\ref{lapla_1_g1} and convolution theorem, yielding
\begin{equation}
\int_{\Gamma}\frac{dq}{2\pi i}e^{qy}\frac{1}{q\sqrt{1+q}}=\int_{0}^{y}dy'~\frac{e^{-y'}}{\sqrt{\pi y'}}=\erf(\sqrt{y})\,.
\end{equation}
Therefore, we obtain
\begin{equation}
g_2(y)=\frac{e^{-y}}{\sqrt{\pi y}}+\erf(\sqrt{y})\,.
\end{equation}
Plugging the expressions for $g_1(y)$ and $g_2(y)$ into Eq.~\ref{convol_F1}, we finally obtain the result in Eq.~\ref{eq:f1z}. Note that this Laplace inversion was also derived in the appendix of Ref.~\cite{MajumdarConvex}. We repeated here the derivation for completeness.

\section{Asymptotic behaviors of $\mathcal{F}(z)$}

\label{app:fl}

To compute the small-$z$ asymptotic behavior of $\mathcal{F}(z)$, we expand the integrand in Eq.~\ref{F} to second order in $z$, yielding
\begin{align}
\mathcal{F}(z)\!  &\approx\!\frac{1}{2}\int_{0}^{\infty}dy\frac{1}{y^6}\sum_{k=0}^{\infty}(-1)^k e^{-\pi^2(2k+1)^2/y^2}\left\{(1+2k)\pi\left[y^2(24-y^2+z^2)-16(2k+1)^2\pi^2\right]\right.\nonumber\\
&\times \left. \left[1-\frac12\left(\frac{\pi(2k+1)z}{2y}\right)^2\right]+4zy\left[y^2-2(2k+1)^2\pi^2\right]\frac{\pi(2k+1)z}{2y}\right\}\,.
\end{align}
Computing the integral over $y$, we find
\begin{equation}
\mathcal{F}(z)\approx \frac{\sqrt{\pi}}{64}(16-z^2)\sum_{k=0}^{\infty} (-1)^{k+1}=\frac{\sqrt{\pi}}{128}(16-z^2)\,,
\end{equation}
where we have used the relation
\begin{equation}
\sum_{k=0}^{\infty} (-1)^{k}=\frac12\,.
\end{equation}
Note that this infinite sum has to be interpreted as the limit
\begin{equation}
\lim_{\alpha\to -1}\left(\sum_{k=0}^{\infty}\alpha^k\right)\,.
\end{equation}

To investigate the limit of large $|z|$, we use the definition of $\mathcal{F}(z)$ and Eq.~\ref{eq:P0l2} to obtain
\begin{equation}
\label{F_int}
\mathcal{F}(z)=\frac{1}{4}\int_{\Gamma}\frac{dq}{2\pi i}e^q \cosh(\frac{\sqrt{q}z}{2})\int_{|z|}^{\infty}dy~\frac{1}{\cosh^3(y\sqrt{q}/2)}\,.
\end{equation}
We perform the change of variable $y\to\tilde{y}=y/|z|$ in Eq.~\ref{F_int}, yielding
\begin{equation}
\mathcal{F}(z)=\frac{|z|}{4}\int_{\Gamma}\frac{dq}{2\pi i}e^q \cosh(\frac{\sqrt{q}z}{2})\int_{1}^{\infty}d\tilde{y}~\frac{1}{\cosh^3(|z|\tilde{y}\sqrt{q}/2)}\,.
\end{equation}
Expanding to leading order for large $z$ and computing the integral over $\tilde{y}$, we find
\begin{equation}
\mathcal{F}(z)\approx\frac23\int_{\Gamma}\frac{dq}{2\pi i}e^q \frac{1}{\sqrt{q}}e^{-z\sqrt{q}}\,.
\end{equation}
Finally, performing the Laplace inversion, we get
\begin{equation}
\mathcal{F}(z)\approx \frac{2}{3\sqrt{\pi}}e^{-z^2/4}\,.
\end{equation}

\section{Moments of $g(z)$}
\label{app:g}
In this appendix, we compute the normalization and the moments of the probability distribution 
\begin{equation}
g(z)=2e^{-z}K_0(2e^{-{z/2}})\,,
\end{equation}
with $-\infty<z<\infty$. The $n-$th moment reads
\begin{equation}
\langle z^n\rangle=\int_{-\infty}^{\infty}\dd z~z^ng(z)=2\int_{-\infty}^{\infty}\dd z~z^ne^{-z}K_0(2e^{-{z/2}})\,.
\end{equation}
Performing the change of variable $z\to w=2 e^{-z/2}$, we obtain
\begin{equation}
\langle z^n\rangle=2^{n}\int_{0}^{\infty}\dd w~\left[\log(\frac{2}{w})\right]^n w K_0(w)\,.
\end{equation}
For $n=0$, we obtain the normalization \cite{Bessel}
\begin{equation}
\int_{-\infty}^{\infty}\dd z~g(z)=\int_{0}^{\infty}\dd w ~w K_0(w)=1\,.
\end{equation}
For $n=1$ and $n=2$ the integral over $w$ can be computed using Mathematica and we obtain
\begin{equation}
\langle z\rangle=2\gamma_E\,,
\end{equation}
where $\gamma_E=0.57721\ldots$ is the Euler constant. Similarly, we find
\begin{equation}
\langle z^2\rangle=4 \gamma_E^2+\frac{\pi^2}{3}\,.
\end{equation}

\end{document}